\documentclass[iop]{emulateapj}

\newcommand{\et}{et\thinspace al.\ }
\newcommand{\arcm}{\ifmmode {' }\else $' $\fi}
\newcommand{\arcs}{\ifmmode {'' }\else $'' $\fi}
\newcommand{\kms}{km\thinspace s$^{-1}$}

\newcommand{\gapp}{$_>\atop{^\sim}$} 
\slugcomment{}

\shortauthors{Young et al.} \shorttitle{GC Systems of 
  NGC~1023, NGC~1055, NGC~7332, NGC~7339}

\begin{document}

\title{Globular Cluster Systems of Spiral and S0 Galaxies: \\Results
  from WIYN Imaging of NGC~1023, NGC~1055, NGC~7332 and NGC~7339}

\author{Michael D. Young, Jessica L. Dowell, and Katherine L. Rhode}
\affil{Department of Astronomy, Indiana University, 727 East Third
  Street, Bloomington, IN 47405-7105; youngmd@indiana.edu,
  jlwind@astro.indiana.edu, rhode@astro.indiana.edu}

\begin{abstract}
\label{section:abstract}
We present results from a study of the globular cluster (GC) systems
of four spiral and S0 galaxies imaged as part of an ongoing wide-field
survey of the GC systems of giant galaxies.  The target galaxies ---
the SB0 galaxy NGC~1023, the SBb galaxy NGC~1055, and an isolated pair
comprised of the Sbc galaxy NGC~7339 and the S0 galaxy NGC~7332 ---
were observed in $BVR$ filters with the WIYN 3.5-m telescope and
Minimosaic camera.  For two of the galaxies, we combined the WIYN
imaging with previously-published data from the Hubble Space Telescope
and the Keck Observatory to help characterize the GC distribution in
the central few kiloparsecs. We determine the radial distribution
(surface density of GCs versus projected radius) of each galaxy's GC
system and use it to calculate the total number of GCs ($N_{GC}$). We
find $N_{GC}$ $=$ $490 \pm 30$, $210 \pm 40$, $175 \pm 15$, and $75
\pm 10$ for NGC~1023, NGC~1055, NGC~7332, and NGC~7339,
respectively. We also calculate the GC specific frequency ($N_{GC}$
normalized by host galaxy luminosity or mass) and find values typical
of those of the other spiral and E/S0 galaxies in the survey. The two
lenticular galaxies have sufficient numbers of GC candidates for us to
perform statistical tests for bimodality in the GC color
distributions.  We find evidence at a high confidence level ($>$95\%)
for two populations in the $B-R$ distribution of the GC system of
NGC~1023.  We find weaker evidence for bimodality ($>$81\% confidence)
in the GC color distribution of NGC~7332.  Finally, we identify eight
GC candidates that may be associated with the Magellanic dwarf galaxy
NGC~1023A, a satellite of NGC~1023.
\end{abstract}


\section{Introduction}
\label{section:introduction}

Studies of the spatial distribution, metallicities, and kinematics of
the globular clusters (GCs) in the Milky Way have provided us with
fundamental insights into the formation and evolutionary history of
our own Galaxy (e.g., Searle \& Zinn 1978, Zinn 1985, Armandroff \&
Zinn 1988, Zinn 1993, Mackey \& van den Bergh 2005).  This link exists
more generally as well: studies of GCs in other giant galaxies can
provide us with clues to the host galaxies' origins and structure.
These luminous, old star clusters seem to mark the major star
formation and merger episodes in galaxies (e.g., Schweizer 1987,
Whitmore et al. 1993, Whitmore \& Schweizer 1995, Brodie \& Strader
2006 and references therein) as well as acting as kinematic and
dynamical tracers of the outer regions of galaxies, which are
difficult to probe via integrated light (e.g., Zepf et al.\ 2000, C\^ot\'e
et al.\ 2001, Richtler et al. 2004, Bridges et al. 2007, Strader et
al. 2011).
%

Steady progress has been made in our understanding of extragalactic GC
systems in the $\sim$20 years since Harris' 1991 comprehensive annual
review of GC systems in galaxies beyond the Local Group.
Many of the major advances in extragalactic GC system research were
made possible by the Hubble Space Telescope (HST): for example, HST
has enabled systematic surveys of the GC systems of hundreds of
galaxies in the (relatively nearby) Virgo and Fornax clusters (C\^ot\'e et
al. 2004, Jord\'an et al. 2007) as well as detection of GCs in galaxy
clusters $>$100~Mpc away (e.g., West et al. 2011).
Nevertheless one of the deficiencies pointed out in both the book on
globular cluster systems by Ashman \& Zepf (1998) and the Annual
Review by Brodie \& Strader (2006) is the relative rarity of
wide-field CCD studies that are able to measure accurate global
properties of giant galaxy GC populations.  Reliable measurements of
quantities like the spatial distribution of the GC system, the total
number of GCs, the GC specific frequency, and the global color
distribution require wide-field CCD imaging in multiple filters.  A
CCD camera with a field-of-view (FOV) many arc~minutes on a side is
needed to cover the full radial extent of the GC systems of galaxies
within $\sim$20$-$30~Mpc and accurate multi-color photometry
is crucial for minimizing contamination from non-GCs (e.g., Rhode \&
Zepf 2001, Dirsch et al.\ 2003).

Our group is engaged in an ongoing survey aimed at quantifying the
global properties of the GC populations of giant galaxies.  We use
multi-color imaging with mosaic CCD cameras and target giant spiral,
S0, and elliptical galaxies at distances between $\sim$10$-$25~Mpc.
To date we have observed $>$30 giant galaxies with a range of masses,
morphological types, and luminosities; results for the galaxies we
have finished analyzing are published in a series of papers (Rhode \&
Zepf 2001, 2003, 2004; Rhode et al.\ 2005, 2007, 2010, Hargis et
al.\ 2011; hereafter RZ01, RZ03, RZ04, R05, R07, R10, H11).  Other
groups who use multi-color wide-field imaging techniques have for the
most part targeted massive elliptical galaxies with very populous,
extended GC systems (e.g., Dirsch et al.\ 2003, imaging of NGC~1399;
Tamura et al.\ 2006, imaging of M87; Dirsch et al.\ 2005, imaging of
NGC~4636).  We have recently observed a number of moderate-luminosity
spiral and S0 galaxies in order to fill in specific gaps in our survey
and in the general census of giant galaxies that have been imaged with
wide-field CCD cameras.
Furthermore, these galaxies are a good match, in terms of their
observational requirements, for the FOV and sensitivity of the WIYN
3.5-m telescope and Minimosaic imager.

In this paper we present results from WIYN Minimosaic imaging of the
GC populations of four moderate-luminosity spiral and lenticular
galaxies: the SB0 galaxy NGC~1023, the SBb galaxy NGC~1055, and the
isolated S0/Sbc galaxy pair NGC~7332 and NGC~7339.  Basic properties
of the four galaxies --- morphological type, distance, absolute
magnitude, stellar mass, and environment --- are listed in
Table~\ref{table:galstats}.
The GC systems of two of the galaxies, NGC~1023 and NGC~7332, were
studied by \citet{lb00} and \citet{forbes01}, respectively, so we have
combined their previous results with our wide-field data to produce
the full GC system radial distributions for those two galaxies.  Also
included in our WIYN imaging of NGC~1023 is the Magellanic dwarf
galaxy NGC~1023A, so we identify a subsample of GC candidates that may
be associated with NGC~1023A and estimate this galaxy's total GC
population and specific frequency.

The paper is laid out as follows.  The next two sections 
describe the data acquisition and image reductions
(Section~\ref{section:reductions}) and the detection 
and selection of GC candidates around each galaxy
(Section~\ref{section:detection}).
Section \ref{section:analysis2} addresses the issues of completeness
and contamination in the GC candidate samples.  The results of the
analysis --- including the radial profiles of each galaxy's GC system,
the total number and specific frequency of GCs, and analysis of the GC
color distributions --- are presented in
Section~\ref{section:results}. The last section of the paper gives a
brief summary of the main conclusions.

\section{Observations \& Image Reductions}
\label{section:reductions}
\begin{figure}
\plotone{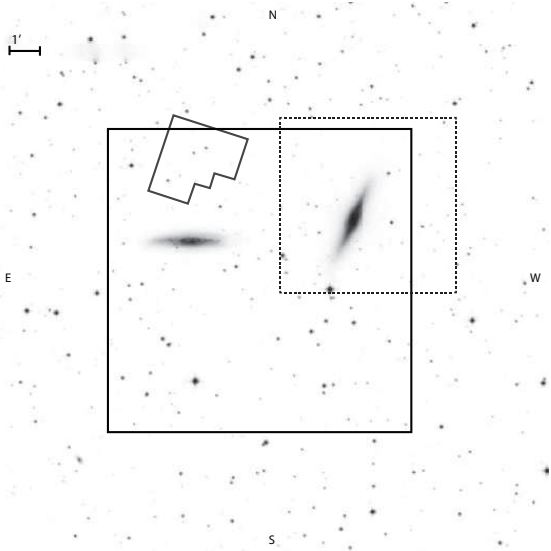}
\caption{ Digitized Sky Survey image of the S0 peculiar
  galaxy NGC~7332 (right) and the Sbc spiral galaxy NGC~7339 (left)
  with the location of the WIYN Minimosaic pointing indicated by a large box.
  The HST WFPC2 pointing discussed in Section~\ref{section:hst1} is
  marked just north of NGC~7339.  The box marked with a dotted
  line around NGC~7332 marks the FOV of the Keck observation analyzed
  by \citet{forbes01} and discussed in Section~\ref{section:hst
    data}. }
\label{fig:n7332 finder}
\end{figure}

Images of the target galaxies were obtained in September 2008 and
September 2009 at Kitt Peak National Observatory with the 3.5~m WIYN
telescope\footnote{The WIYN Observatory is a joint facility of The
  University of Wisconsin-Madison, Indiana University, Yale
  University, and the National Optical Astronomy Observatory (NOAO).}
and Minimosaic camera. The detector is comprised of two 2048 $\times$
4096 CCDs with 0.141\arcs\ pixels, and a FOV of 9.6\arcm\ $\times$
9.6\arcm\ when paired with the WIYN telescope.  The relatively large
FOV allowed us to image the isolated galaxy pair NGC~7332 and NGC~7339
simultaneously.  To obtain sufficient radial coverage of these
galaxies' GC systems, NGC~7339 was positioned near the top center of
the field, while NGC~7332 was placed in the lower right. NGC~1023 and
1055 were both positioned in the center of one CCD (away from a narrow
gap between the two Minimosaic CCDs).  Figures~\ref{fig:n1023
  finder}-\ref{fig:n7332 finder} show the WIYN FOV for each target
galaxy overlaid on a Digitized Sky Survey (DSS)~\footnote{The
  Digitized Sky Surveys were produced at the Space Telescope Science
  Institute under U.S. Government grant NAG W-2166. The images of
  these surveys are based on photographic data obtained using the
  Oschin Schmidt Telescope on Palomar Mountain and the UK Schmidt
  Telescope. The plates were processed into the present compressed
  digital form with the permission of these institutions.}  image.  To
help us differentiate GCs from contaminating foreground stars and
background galaxies, multiple exposures were taken in each of three
broadband filters ($BVR$).  For a list of exposure times see
Table~\ref{table:observations}.

\begin{figure}
\plotone{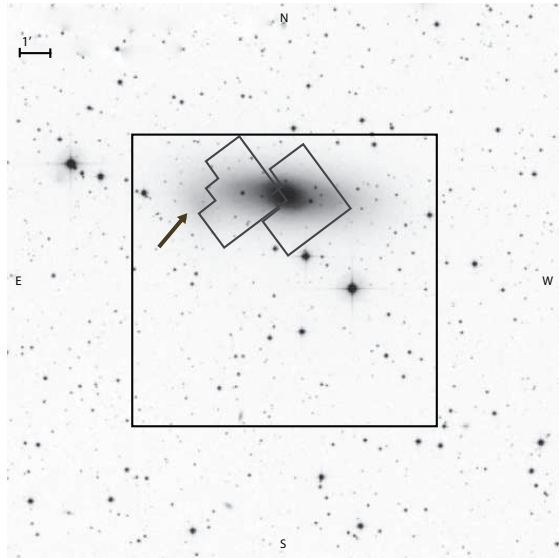}
\caption{Digitized Sky Survey image of the S0 galaxy NGC~1023 with the
  location of the WIYN Minimosaic pointing indicated by the large box.
  The locations of two HST WFPC2 pointings that were first analyzed by
  \citet{lb00} and which are discussed in Section~\ref{section:hst
    data} are overlaid on the central portion of the galaxy. The
  location of the dwarf galaxy NGC~1023A (discussed in
  Section~\ref{section:dwarf}) is marked with an arrow.}
\label{fig:n1023 finder}
\end{figure}

\begin{figure}
\plotone{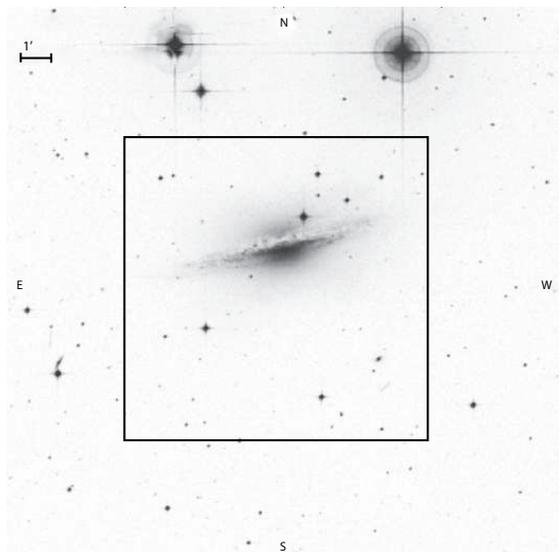}
\caption{Digitized Sky Survey image of the Sb spiral galaxy NGC~1055
  with the location of the WIYN Minimosaic pointing indicated by a large box.}

\label{fig:n1055 finder}
\end{figure}

Observations of photometric standard star fields \citep{land92} were
made in photometric conditions throughout the first night of the 2008
observing run and the third night of the 2009 run, along with at least
one exposure of each target galaxy in all three filters.  These
observations were used to determine the color terms and zero point
constants for the data taken on the photometric nights.  The
calculated errors on the zero point constants were all less than or
equal to 0.007 magnitude, which demonstrates that those nights were
genuinely photometric.

Standard reductions (overscan and bias level subtraction, flat-field
division) were performed on all of the Minimosaic images using tasks
in the IRAF package MSCRED.  The MSCRED tasks {\it msczero},
{\it msccmatch}, and {\it mscimage} were then used to convert
the multi-extension FITS images into single-extension images.  The
images taken of a given target galaxy were aligned to each other. A
constant sky background level was measured for each image and
subtracted.  Images of a given galaxy taken in the same filter were
then scaled to a common flux level and combined using the IRAF task
{\it imcombine} and the {\it ccdclip} pixel rejection algorithm.
The constant sky background level was then added back to the final
combined image.  The result of this processing was a single deep,
stacked image of each target galaxy in each filter.  The mean
full-width at half-maximum of the point-spread function (FWHM PSF) in
the combined $BVR$ images ranged from $0.6\arcs-1.0\arcs$ for
NGC~1023, $0.9\arcs-1.0\arcs$ for NGC~7332 \& NGC~7339, and
$1.0\arcs-1.1\arcs$ for NGC~1055.  The method we used for calibrating
the photometry in the final stacked images (using the data taken on
the photometric nights of the 2008 and 2009 runs) is explained in
Section~\ref{section:photometry}.


\section{Detection and Analysis of the Globular Cluster Systems}
\label{section:detection}

\subsection{Source Detection}
\label{section:source detection}

To facilitate detection of GC candidates close to each target galaxy,
we executed a series of steps to remove the diffuse galaxy light from
the deep, stacked images. We began by smoothing each of the images
with a circular median filter with a diameter 7$-$7.5 times the mean
FWHM PSF of the image.  The size of the median filter was chosen after
testing to find the optimal size for removing diffuse light from the
galaxy while preserving the PSF of the compact, star-like objects
around the galaxy.  This smoothed image was then subtracted from the
original version.  A constant background level was added to the
resultant images to preserve the photometric integrity of later
measurements. The rest of the analysis steps were performed on these
``galaxy-subtracted'' images (one such image for each filter and each
galaxy field).

Some regions of the galaxy-subtracted images --- such as small areas
around saturated stars and high-background areas in the galaxy disks
--- were masked out prior to the detection step in order to minimize
spurious detections.  The IRAF task {\it daofind} was then used to
identify sources above a given signal-to-noise threshold in each
image.  The source lists for the $B$, $V$, and $R$ filters were
matched, and only sources that appeared in all three filters were
retained in the final list of detected objects for a particular galaxy
field.  For NGC~7332 \& NGC~7339 a total of 1126 objects appeared in
all three filters, while NGC~1055 had 343 matched objects and NGC~1023
had 2141 matched objects.  
The relatively small numbers of objects found in the NGC~1055 field
compared to the other two fields had two causes.  First, NGC~1055 has
a higher Galactic latitude than the other galaxy fields --- its
latitude is $b$ $=$ $-$51~degrees, compared to $b$ $=$ $-$19~degrees
for NGC~1023 and $b$ $=$ $-$29 degrees for NGC~7332 --- so many fewer
Galactic foreground stars appear in the NGC~1055 field. Secondly,
NGC~1055 is an Sb spiral galaxy, whereas NGC~1023 and NGC~7332 are S0
galaxies; we are finding in our GC system survey that spiral galaxies
typically have many fewer GC candidates around them than early-type
galaxies.

\subsection{Removing Extended Sources}
\label{section:ext source cut}

\begin{figure}
\plotone{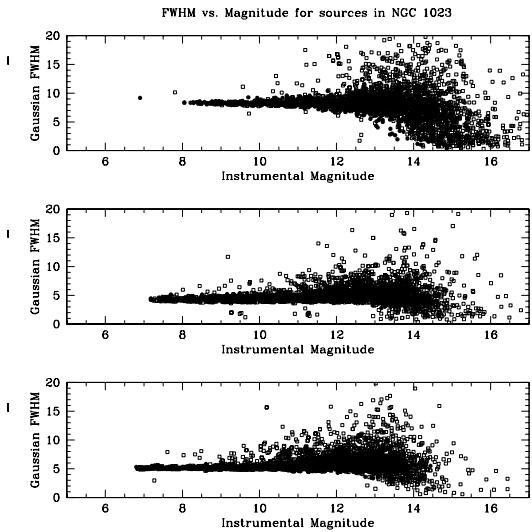}
\caption{ An example of the extended source cut step: the
  plots show measured FWHM vs instrumental $B$, $V$, and $R$ magnitude
  in the final combined images of NGC~1023.  Filled circles mark the
  917 objects that passed the extended source cut in all three
  filters.  Open squares mark rejected objects.  Similar cuts were
  applied to the source lists for all target galaxies.}
\label{fig:1023magvfwhm}
\end{figure}

Given the distances to the target galaxies ($\sim$10$-$20~Mpc; see
Table~\ref{table:galstats}), GCs should appear as unresolved point
sources in our ground-based images. The median half-light radius for
Milky Way GCs is 3~pc \citep{az98}, which subtends 0.06\arcs\ at 10~Mpc, and
our images have FWHM PSF values of 0.6$-$1.1\arcs.
Thus any extended objects are not real GCs and need to be removed from
the source lists. To accomplish this, we plotted the FWHM of the
detected objects against instrumental magnitude (separately for each
galaxy field and for each filter).  In such a diagram, point sources
produce a tight grouping that flares at faint magnitudes.  Sources
that deviated markedly from this sequence
in any of the three filters were eliminated from the source list.  Our
extended source cut for NGC~1023 is illustrated in Figure
\ref{fig:1023magvfwhm}; a similar cut was applied to all three target
fields.  Once the extended objects were removed, the source list for
NGC~1023 contained 917 objects.  The source list for NGC~1055 included
257 objects, and the list for NGC~7332/NGC~7339 had 642 objects.

\subsection{Aperture Photometry of Matched Point Sources}
\label{section:photometry}

Calibrated $B$, $V$, and $R$ magnitudes for the matched list of point
sources in each galaxy field were derived using the following
procedure.  Aperture photometry was performed on the point sources in
each field (i.e., the 917 sources in the NGC~1023 field, the 257
objects in the NGC~1055 field, and the 642 objects in the
NGC~7332/NGC~7339 field) using the IRAF task PHOT.  The radii of the
apertures used to do the photometry were set equal to the mean FWHM
value of the image.  An aperture correction was derived for each image
by calculating the mean of the difference between the total magnitude
and the magnitude measured within a one-FWHM-radius aperture for a set
of $\sim$10$-$20 bright point sources.

In addition to the aperture correction, a bootstrap correction was
determined for calibrating our stacked images, which are a combination
of images taken on both photometric and non-photometric nights.  To
calculate the bootstrap correction, instrumental magnitudes of bright
objects were compared between single-frame images taken on photometric
nights and the final stacked images of a given target field.  The mean
magnitude difference between the single photometric image and the
final, combined image was calculated for each case.
We also determined foreground Galactic extinction corrections (in the
direction of each target galaxy field) from the maps of infrared dust
emission presented in \citet{schlegel98}.

Total instrumental magnitudes were calculated for the matched point
sources by applying the appropriate aperture, bootstrap, and airmass
corrections to the instrumental magnitudes measured with the
one-FWHM-radius aperture.  These magnitudes were then calibrated using
the appropriate photometric coefficients and Galactic extinction
corrections, to produce final $B_0$, $V_0$, and $R_0$ magnitudes and
colors for the point sources.

\subsection{Final Selection of GC Candidates}
\label{section:final selection}

In the final GC candidate selection step, we establish an expected
range of $BVR$ magnitudes and colors for GCs at the target galaxies'
distances and keep only the objects that are consistent with those
values.  The GC luminosity function (GCLF) of the Milky Way and other
giant galaxies indicates that the most luminous GCs will typically
have $M_V$ $\sim$ $-$11 \citep{az98}.  For NGC~1055, we eliminated
point sources with $M_V$ brighter than $-$11 (assuming the distance
modulus given in Table \ref{table:galstats}).  For NGC~7332 and
NGC~7339, there were three point sources that were close to the host
galaxies, looked like GCs, and had $M_V$ values a few tenths of a
magnitude brighter than $-$11.  We relaxed the bright-end magnitude
criterion slightly in order to include only these three specific
objects.
In the case of NGC~1023, we adjusted our bright-end cut to include a
luminous, spectroscopically-confirmed GC originally identified by
\citet{lb00}.  Relaxing the bright-magnitude criterion yielded an
extra seven objects around NGC~1023 that have GC-like colors and $M_V$
magnitudes brighter than $-$11. The sequence numbers, positions, $V$
magnitudes, and $B-V$ and $V-R$ colors and corresponding errors for
these seven objects are given in Table~\ref{table:n1023 bright
  objects}.  Given our assumed distance modulus for NGC~1023
($m-M$$=$30.29), the $M_V$ values for these seven objects range from
$-$11.4 to $-$11.0.  (We note that two of the objects, 119 and 133,
are located several arc~minutes away from the galaxy center and thus
would seem less likely than the objects at small projected radial
distances to be bona fide GCs.)  For the NGC~7332/7339 field we also
implemented a faint source cut, eliminating objects with $V > 24.4$.
We did this to remove a number of faint objects that had large
magnitude errors (\gapp0.15) and were located many arc~minutes from
the target galaxy; these are likely to be unresolved background
galaxies.  For similar reasons a faint cut of $V > 23.5$ was
implemented for both the NGC~1023 and NGC~1055 fields.  (Note that any
GCs we miss due to the magnitude incompleteness of our images and the
cuts imposed on the samples at the faint end are accounted for in the
GCLF correction step described in Section~\ref{section:gclf}.)

We also implemented the standard $BVR$ color selection that we use for
the wide-field survey (see \citealp{rz01} for more details): we
removed point sources that lie more than $3\sigma$ away from the mean
$V-R$ vs.\ $B-V$ relation for Galactic GCs with metallicities between
[Fe/H] of $-2.5$ and $0.0$.  When we test whether each source is
consistent within $3\sigma$ of the Milky Way GC color-color relation,
we take into account the object's photometric error.  For NGC~7332 we
added a single object that fell just barely outside the color criteria
and was located close ( $< 0.4$\arcm) to the galaxy center.

GC candidates positioned very close to the host galaxy disk may be
subject to additional reddening and thus might appear in the $BVR$
color-color diagram just redward of the GC candidate selection box.
Point sources that appeared in this part of the color-color diagram
were individually examined in the images.  For all four of the target
galaxies, none of the objects just redward of the color selection box
were positioned close to the host galaxy disk in the images, so we did
not add any such candidates to our final list.

An additional complicating factor in the detection of GC candidates
around NGC~1023 was the presence of the dwarf galaxy NGC~1023A.  The
galaxy is classified as a Magellanic irregular in RC3 \citep{devauc91}
and is located $\sim$2.5\arcm\ (a projected distance of $\sim$8~kpc)
to the southeast of NGC~1023 in the image. According to the NASA
Extragalactic Database (NED), NGC~1023A is a member of the NGC~1023
group.  We determined that the light from NGC~1023A extends over a
circular region with a radius of 0.6\arcm\ in the $V$-band image.
We did this by plotting the pixel intensities along the radial
direction between NGC~1023A and NGC~1023 to find the radial distance
at which the light from NGC~1023A begins to dominate.  We measured
this radial distance in all three of the combined images and averaged
the results to come up with the radial size of NGC~1023A in the
images.
We excluded this region from the rest of our analysis of NGC~1023's GC
system.  Eight of the point sources with magnitudes and colors like
GCs are located within that circular region.  We note that, without
additional information, we cannot definitively say with which galaxy
these objects are associated. We explore the properties of these eight
objects in further depth in Section \ref{section:dwarf}, including
estimating how many of them are likely to be GCs in the dwarf galaxy
(rather than part of NGC~1023's GC system).  We correct for the area
that was masked around NGC~1023A when we construct the radial profile
for NGC~1023's GC system (see Section \ref{section:radial profile}).

\begin{figure}
\plotone{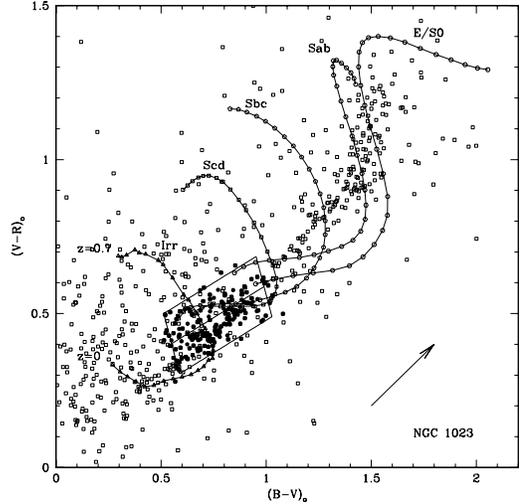}
\caption{ Selection of GC candidates in NGC~1023 based on
  $V-R$ and $B-V$ colors and $V$ magnitudes. The 917 point sources
  that survived the extended source cut in all three filters are
  marked with open squares.  The final selected 192 GC candidates are
  indicated with filled circles.  The box marks the area of our color
  selection.  Sources outside the color selection area, but with
  colors and corresponding errors consistent with the selection
  criteria are retained as GC candidates.  A reddening vector
  equivalent to $A_V$ = $1$ is in the lower right-hand corner of the
  image.  The curves mark the colors of galaxies of several
  morphological types in a redshift range of z = 0 to z = 0.7,
  indicating possible sources of background contamination (see Rhode
  \& Zepf 2001 for details of how the curves were generated).}
\label{fig:1023 color-color}
\end{figure}

The final GC candidate list for NGC~1023 contained 192 objects, while
NGC~1055 had 60 GC candidates and NGC~7732 and NGC~7339 together had
105 GC candidates. Figures \ref{fig:1023 color-color} --- \ref{fig:7332
  color-color} show the $BVR$ color-color diagrams for the point
sources detected around the four galaxies.  Open squares represent the
detected point sources, and the filled circles show the objects that
were selected for the final GC candidate list.  The GC color selection
box and the mean $V-R$ vs.\ $B-V$ relation for Milky Way GCs (the line
marking the midpoint of the box) are marked with solid lines.  The
tracks indicate the expected colors for galaxies of different
morphological types at redshifts between $z=0$ and $z=0.7$, to
illustrate possible sources of background galaxy contamination; i.e.,
to show what types of galaxies, and at what redshifts, are likely to
lie in the same part of the $BVR$ color-color plane as GCs.  (See RZ01
for a full discussion of how the color cut for the survey was
developed and details about how these galaxy tracks were produced.)

\begin{figure}
\plotone{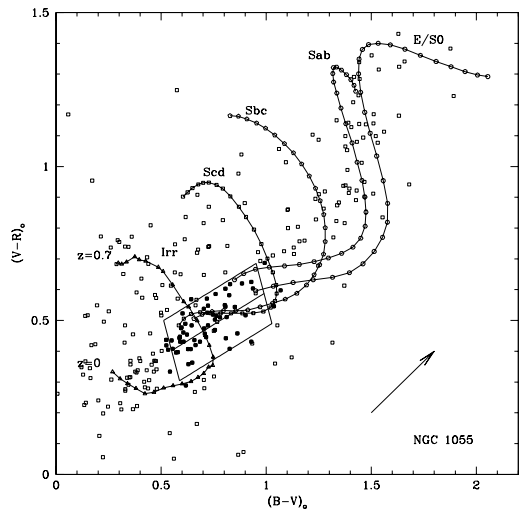}
\caption{ Selection of GC candidates in NGC~1055 based on
  $V-R$ and $B-V$ colors and $V$ magnitudes, plotted in the same way
    as in Figure~\ref{fig:1023 color-color}. The 257 point sources
  that passed the extended source cut in all three filters are
  marked with open squares; the 60 GC candidates are marked with open
  circles.
}
\label{fig:1055 color-color}
\end{figure}

\begin{figure}
\plotone{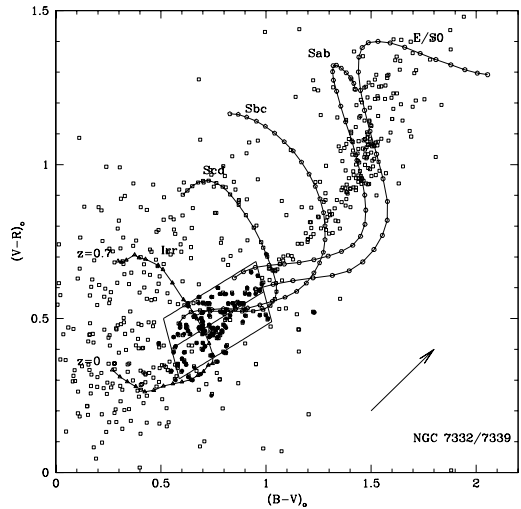}
\caption{ Selection of GC candidates in both NGC~7332 and NGC~7339
  based on $V-R$ and $B-V$ colors and $V$ magnitudes, plotted in the
  same way as in Figure~\ref{fig:1023 color-color}. These galaxies
  were included in the same image so the GC selection was done over
  the full field, for both galaxies at once.  The 642 point sources
  that survived the extended source cut in all three filters are
  marked with open squares; the 105 GC candidates in the
  NGC~7332/NGC~7339 field are marked with filled circles.}
\label{fig:7332 color-color}
\end{figure}

\section{Additional Analysis Steps}
\label{section:analysis2}

\subsection{Completeness Testing and Detection Limits}
\label{section:completeness}

In order to determine the limit of our ability to detect point sources
of a given magnitude in the galaxy-subtracted images, we created a
series of tests to quantify the completeness level as a function of
object magnitude in each filter for each galaxy field. We first
measured the typical PSF for each image.  We added 200 artificial
point sources with the appropriate PSF and a given magnitude to each
of the combined images ($B$, $V$, and $R$).  We did this in
0.2-magnitude steps over a range of 6$-$7 magnitudes, creating a
separate image for each magnitude step (and each filter).  The images
with the artificial point sources were then processed in a series of
steps identical to the detection steps we had executed to find GC
candidates in the original images.  For each magnitude step and each
filter, we determined what fraction of the artificial sources were
recovered. This gave us an estimate of the completeness at that
magnitude, i.e., our ability to detect point sources of a given
magnitude in a given image. The result was a series of completeness
curves, one for each filter, for each set of galaxy-subtracted
combined images.  The completeness information was taken into account
in our analysis of what fraction of the GC luminosity function we had
covered for each galaxy (see Section~\ref{section:gclf}).  The 50\%
completeness limits are typically $\sim24$-$25$ magnitudes; Table
\ref{table:completeness} lists the exact values for each image.

\subsection{Contamination}

Contamination of the GC candidate lists by foreground stars and
background compact galaxies occurs because some of these objects may
match all of the selection criteria for GC candidates.  Here we
present our methods for quantifying the contamination level and, where
possible, correcting for the presence of contaminating objects.

\subsubsection{Model Prediction for Stellar Contamination}
\label{section:star count model}

We used a Galactic stellar population synthesis model to estimate the
level of stellar foreground contamination that might be present in the
GC candidate lists. Hereafter we refer to the model as the Besan\c con
model; a complete description of the model and the observational
constraints used to develop it are given in \cite{rrdp03}.  Briefly,
the model divides the Milky Way Galaxy into four distinct stellar
populations: a thin disk, a thick disk, a spheroid, and a bulge.  Each
component is defined by a star formation rate, initial mass function,
metallicity distribution, kinematics, and evolutionary tracks within
the model.  Given a range of stellar types, magnitudes, and colors,
the model uses Monte Carlo methods to produce Galactic star counts for
a specific region of the sky.  Comparisons of the star count
predictions of the Besan\c con model to photometric surveys have shown
reasonable agreement (e.g., \citealp{rrc00} and \citealp{rr01}).

We used the Besan\c con model to estimate the number of Galactic
foreground stars that would appear within the fields of view of our
Minimosaic pointings.  In order to arrive at an estimate of the number
of possible contaminants in the GC candidate samples, we set the
inputs of the model to provide counts only for Galactic stars with
$B-V$ colors, $V-R$ colors, and $V$ magnitudes defined by the range of
properties in the final selected GC candidate list for each galaxy.
To account for the statistical nature of the model, we ran it ten
times per target galaxy field and then calculated the mean Galactic
star counts in each case.  We then used this number and the FOV of the
Minimosaic images to compute the surface density of contaminating
Galactic stars for the galaxy field.  The contamination was estimated
to be 0.15~arcmin$^{-2}$ for NGC~1055.  NGC~1023 has an estimated
foreground contamination of 0.56~arcmin$^{-2}$, while the field for
NGC~7332 and 7339 has an estimated foreground contamination of
0.54~arcmin$^{-2}$.  As noted in the discussion of source detection
(Section~\ref{section:detection}), the relatively low level of
foreground stellar contamination for NGC~1055 compared to the other
target galaxies is expected because of its relatively high Galactic
latitude.

\subsubsection{Examining the WIYN GC Candidates in Archival HST Images}
\label{section:hst1}

Another way to evaluate the level of contamination in the WIYN data is
by examining archival HST images~\footnote{Based on observations made
  with the NASA/ESA Hubble Space Telescope, and obtained from the
  Hubble Legacy Archive, which is a collaboration between the Space
  Telescope Science Institute (STScI/NASA), the Space Telescope
  European Coordinating Facility (ST-ECF/ESA) and the Canadian
  Astronomy Data Centre (CADC/NRC/CSA).}.  Many compact background
galaxies that
appear as point sources in ground-based imaging may be resolved by
HST, so determining whether some of the WIYN GC candidates are
actually extended objects can help us assess how much background galaxy
contamination is present in the WIYN GC candidate lists.

Accordingly, we analyzed HST archival data for NGC~7332 \& NGC~7339 to
determine if any of the WIYN GC candidates were background
galaxies. We found one Wide Field Planetary Camera 2 (WFPC2) image in
the Hubble Legacy Archive (HLA) that was taken in an appropriate
broadband filter (F606W) and had sufficient depth (a combination of
three exposures totaling 1500 seconds) for detecting our GC
candidates. The WFPC2 pointing (HST proposal ID 7566, PI: Green) was
$\sim$2.5\arcm\ from the center of NGC~7339.  The footprint of the
WFPC2 pointing is marked with a solid line in Figure~\ref{fig:n7332
  finder}.  Five of the WIYN GC candidates lie within the WFPC2 FOV.
Aperture photometry was performed on these sources in the HST images,
using apertures with radii of 0.5 and 3.0 pixels. The ratio of counts
in the larger aperture to the smaller aperture is expected to be $<$
12 for compact objects on the PC chip and $<$ 8 for compact objects on
the WF chip \citep{kundu99}.  All five of the WIYN GC candidates had
count ratios in the WFPC2 images consistent with their being point
sources.  It is important to note here that this is not necessarily
indicative of zero background contamination from background galaxies
(since only five of the WIYN GC candidates appear in the WFPC2
pointing), but it may indicate at least that the WIYN candidate sample
is unlikely to be dominated by background objects.

For NGC~1023, there were four archival images of sufficient depth
(each a combination of two exposures totaling 2400 seconds) and taken
in useful filters (F555W and F814W) in the HLA that we could use to
assess our background galaxy contamination level.  The HST images
included two slightly overlapping WFPC2 pointings near the galaxy
center from HST proposal ID 6554 (PI: Brodie). The WFPC2 footprints are
marked in Figure~\ref{fig:n1023 finder}. Twenty of our WIYN GC
candidates were identified in both filters, and of those, two had
count ratios indicating they were extended objects. This again seems
to suggest that the WIYN GC candidate samples are not dominated by
contamination from background galaxies.  We did not remove these
objects from our candidate list at this stage, since the presence of
contaminants was dealt with by the application of the overall
contamination correction (Section \ref{section:asymptotic}).  Manually
removing only a select few contaminating objects at this point in the
analysis would effectively result in a double correction and an
incorrect result for the derived GC system radial profile.  These two
objects were, however, excluded during the color distribution (Section
\ref{section:tblue}) analysis step, where the measured properties of
individual objects plays a role.

There are currently no images of NGC~1055 in the HST archive.  Thus we
could not do this type of analysis for the NGC~1055 field and GC
candidate list.

\subsubsection{Contamination Correction Based on the Asymptotic
Behavior of the Radial Profile}
\label{section:asymptotic}

Since the GC population is expected to fall off with increasing
distance from the center of the galaxy, we assume that beyond some
point the GC candidate sample will be dominated by contaminating
objects (compact objects with magnitudes and colors like GCs, which
are actually stars or background galaxies).  At large projected radii
the radial distribution of GC candidates should flatten to some
constant value; the surface density in the outermost portion of the
radial profile thus provides us with an estimate of the surface
density of contaminants.
We refer to this estimate as the ``asymptotic contamination
correction''.

To investigate the radial distribution of each galaxy's GC system as
well as to determine the asymptotic contamination correction, we
assigned the GC candidates to one of a series of circular annuli
extending outward from the galaxy center. The annuli sizes we used to
create this initial radial profile for each galaxy varied between 0.5'
wide to 0.8' wide, depending on how many GC candidates appeared around
the galaxy and how the candidates were distributed.  For each annulus,
we calculated a corresponding ``effective area'', equal to the area of
the annulus minus any area in which GC candidates could not be
detected (e.g., masked or saturated regions, or parts of the annuli
that extended off the edges of the images).  We then calculated the
surface density of GC candidates (number per effective area) in each
annulus, and plotted this quantity versus the mean projected radius of
the corresponding annuli.

For the GC system of NGC~1055, we found that the surface density of GC
candidates converged to a fairly constant value and remained flat in
the annuli with mean radii greater than 4.7\arcm.  Taking the weighted
mean of the surface density in the annuli beyond this point, we find a
contamination value of $0.30 \pm 0.13$ objects~arcmin$^{-2}$. For
NGC~1023, the profile converged to a relatively constant level at
radii beyond 6.7\arcm; the weighted mean surface density in these
outer annuli is $0.74 \pm 0.20$ objects~arcmin$^{-2}$.

NGC~7332 \& NGC~7339 appear on the sky near to each other ($<$
6\arcm\ separation), although they are not thought to be interacting
\citep{devauc76}.  The proximity of these galaxies in the Minimosaic
pointing complicated our efforts to determine an asymptotic
contamination correction.  Both galaxies showed a GC surface density
that decreased with increasing radius in the first few annuli of the
profile, but then increased again due to the presence of the
neighboring galaxy's GC system (i.e., the GC surface density around
NGC~7332 dropped with radius in the first few arc~minutes around the
galaxy, but then increased in the region around NGC~7339, and vice
versa).  To remedy this, we noted the approximate radius at which the
GC population around each galaxy fell to a constant level, and then
placed a circular mask with that radius over the galaxy and its GC
system.  For both galaxies, a circular region with radius
1.8\arcm\ ($\sim$12~kpc) was sufficient to mask out the GC population.
We used the image with both masks in place to determine the asymptotic
contamination correction for the entire field.  Because NGC~7332 and
NGC~7339 appear in the same WIYN pointing and we used a single set of
GC candidate selection techniques for the $BVR$ images of that field,
the surface density of contaminants that we apply should be the same
for the two galaxies.  Combining their asymptotic contamination
corrections yields a weighted average contamination value of $0.57 \pm
0.07$ objects~arcmin$^{-2}$.

The last step in this process was to take the appropriate asymptotic
contamination correction for each galaxy and multiply it by the
effective area of each annulus in the radial profiles.  This provides
us with an estimate of the number of contaminating objects in each
radial bin of each galaxies' GC system profile. The {\it fraction} of
contaminating objects in each annulus is the number of contaminating
objects divided by the number of GC candidates.  This provides us with
a radially-dependent contamination correction that we can use in
subsequent analysis steps.

We note that for all of the galaxies, the contamination level
determined from the asymptotic behavior of the radial profile is
larger than the level calculated from the Galactic star count models.
This is a useful consistency check; we expect that the contamination
in our GC candidate samples will come from both compact background
galaxies and foreground Galactic stars, so the total number density of
contaminants should be larger than the number density from the
Galactic stellar population models (in the limit that the model
results are accurate).

\subsection{Determining the GCLF Coverage of the WIYN Data}
\label{section:gclf}

We used the list of GC candidates for each galaxy to create a
luminosity function (LF) --- a histogram of the number of GC
candidates in a series of $V$ magnitude intervals --- for that
galaxy's GC system.  The radially-dependent contamination correction
described in Section \ref{section:asymptotic} provides us with an
estimated contamination fraction for each annulus in the radial
profile, so we used it to correct the LF data.  If a GC candidate was
located, for example, in an annulus that had a contamination fraction
of 15\%, then 0.85 was added to the number of objects in the
appropriate $V$ magnitude bin of the LF histogram.  We then divided
each magnitude bin by its completeness correction
(Section~\ref{section:completeness}), yielding a final corrected
GCLF. Because we require that the GC candidates must be detected in
the $B$, $V$, and $R$ images, the completeness correction we apply
takes into account the detection limits in all three filters
(specifically, by convolving the completeness level in each filter at
each $V$ magnitude bin; see RZ01 for additional details). 

\begin{figure}
\plotone{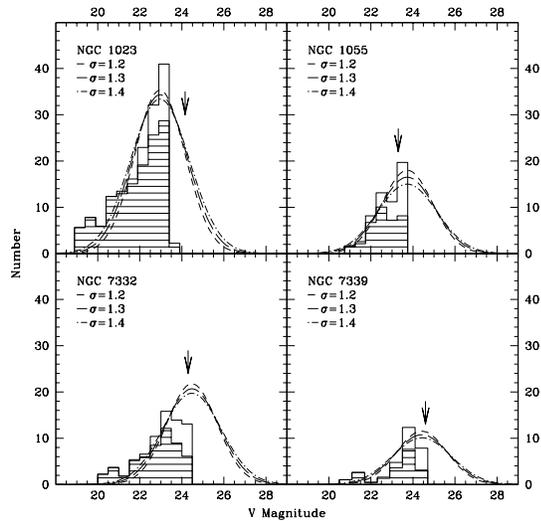}
\caption{GCLF fitting results for the four target galaxies. The shaded
  histogram shows the observed luminosity function of the GC
  candidates in each galaxy, corrected for contamination as described
  in Section~\ref{section:gclf}. The completeness-corrected luminosity
  function that was used for the fitting is marked with a solid
  line. The arrow marks the $V$ magnitude at which the convolved $BVR$
  completeness level is 50\%. (Note that in the case of NGC~1023, the
  faint-magnitude cut applied to the GC candidate list was at $V$ $=$
  23.5, which is brighter than the 50\% completeness threshold.)  The
  best-fitting Gaussian GCLF functions are shown for the three
  different dispersions that were used in the fitting process.}
\label{fig:gclf}
\end{figure}

We cannot reliably determine the exact shape of the GCLF for these
galaxies based on our data, because either we have insufficient
numbers of GC candidates in the samples or we do not follow the
luminosity function far enough past the peak of the GCLF, or
both. Therefore to fit the GCLF and determine how many GCs we may have
missed, we assume that the intrinsic GC luminosity function is a
Gaussian function similar to that of the Milky Way GC system, which
has a peak magnitude at $M_V = -7.33$ and a dispersion of 1.23
\citep{az98}.  Given the distances in Table~\ref{table:galstats}, we
calculated the peak apparent magnitude for the GCLF of our galaxies,
and fitted Gaussians with that peak and dispersions of 1.2, 1.3, and
1.4 magnitude to the LF data.  We allowed the amplitude of this
theoretical GCLF to vary and calculated the fraction of the total area
of the best-fitting theoretical GCLF that was covered by our observed
LF data.  The size and starting position of the LF magnitude bins were
also varied to investigate the effect on the calculated GCLF coverage;
the resulting differences were included in our error calculations in
Section \ref{section:spec freq}.  The observed GCLF data and
best-fitting Gaussian functions four the four target galaxies are
plotted in Figure~\ref{fig:gclf}. The shaded histogram is the
observed, contamination-corrected luminosity function of the GC
candidates and the solid-line historgram is those same data corrected
for the final convolved $BVR$ completeness.  The best-fitting Gaussian
functions for three different dispersions are shown as indicated in
the figure legend.  Table \ref{table:gclf} lists the mean calculated
GCLF coverage for each galaxy.  For NGC~1055, NGC~7332, and NGC~7339,
we detect about one-third of the GC population given the magnitude
depth and detection limits of our data; for NGC~1023, which has a
distance that is significantly closer than the other three galaxies,
we cover about 60\% of the theoretical GCLF.

\section{Results}
\label{section:results}

\subsection{Radial Distributions of the GC Systems}
\label{section:radial profile}

As mentioned in the explanation of the asymptotic contamination
correction (Section~\ref{section:asymptotic}), a radial profile for
the GC system of each galaxy was constructed by first assigning each
GC candidate to one of a series of concentric circular annuli starting
from the center of the host galaxy and continuing out to the edges of
the images. After some experimentation with various radial profile bin
sizes, we settled on annuli corresponding to approximately 3~kpc in
physical size at the distance of each galaxy. Accordingly, the
circular annuli used to create the radial profile for NGC~1023 were
0.8\arcm\ wide, the NGC~1055 annuli were 0.8\arcm\ wide, and for
NGC~7332 0.5\arcm-wide annuli were used.  For NGC~7339 the GC system
is extremely compact around the galaxy, extending to only
$\sim$1\arcm\ from the galaxy center.  This necessitated the use of a
more granular radial bin size (0.3\arcm) to obtain more measurements
at the cost of an increase in the relative error.  Each annulus was
corrected for missing areal coverage due to parts of the annuli being
masked or extending off the edges of the images.

The surface density of contaminants, as determined from the asymptotic
behavior of the radial profile (Section \ref{section:asymptotic}), was
multiplied by the effective area of each concentric annulus to yield
the number of contaminants for that annulus.  After correcting the
total number of GC candidates in each annulus for contamination, we
applied the appropriate GCLF correction, then divided by the effective
area of the annulus to yield the surface density of GCs in each
annulus.  The error on the surface density includes Poisson errors on
the number of GCs and the number of contaminating objects.  We
calculated the mean distance of all unmasked pixels in each annulus,
and used this as the radial distance of each annulus in the radial
profile.  The final, corrected radial profile data for the GC systems
of the four galaxies are listed in Tables \ref{table:1023radprof}
through \ref{table:7339radprof}.  For each projected radial distance,
a surface density and error are listed.  In the case of NGC~1023 and
NGC~7332, we supplemented our WIYN data with data from
previously-published studies of the GC populations to generate
additional points in the GC system radial distributions.  Therefore
the radial profile tables for these galaxies also list the source of
the measurement in the last column. (More details on the supplementary
data are given in the next section.)

The final radial profiles are plotted in Figures \ref{fig:1023 radial
  profile} through \ref{fig:7339 radial profile}.  The top plot in
each figure shows the GC surface density versus projected radius,
while the bottom plot shows the log of the GC surface density versus
the log of the projected radius.  We fitted both a de~Vaucouleurs law
(of the form $\log \sigma_{GC} = a0 + a1 ~r^{1/4}$) and a power law
(of the form $\log \sigma_{GC} = a0 + a1~ \log r$) to each radial
profile. Here, $\sigma_{GC}$ is the surface density of GCs in number
per square arc minute and $r$ is the projected radius in arc~minutes.
The slope, intercept, and reduced $\chi^2$ value corresponding to the
best-fitting power law and de~Vaucouleurs law functions are given in
Table \ref{table:fitting}.  Both functions are shown in the bottom
plots of the radial profile figures
but the function with the lowest reduced $\chi^2$ value was the one
used in subsequent analysis steps (see Section~\ref{section:totalN}).
The power law is the best-fitting function for the radial profiles of
the GC systems of NGC~1055 and NGC~7339.  This is also the case for
NGC~7332, although the data points in the profile (Fig.\ref{fig:7332
  radial profile}) fall below the power law at large radius.  NGC~1023
is best fit by a de~Vaucouleurs profile.  In this case, the points in
the inner radial profile fluctuate between low and high values; the
reasons for this are explored in the next section.

\begin{figure}
\plotone{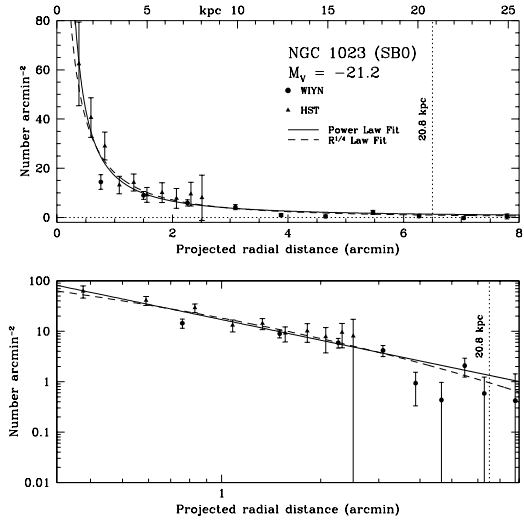}
\caption{ Radial distribution of the GC system of NGC 1023. The top
  plot shows the surface density of GCs as a function of projected
  radial distance ($r$) from the galaxy center.  The bottom plot shows
  the log of the GC surface density versus the log of the projected
  radius.  The solid line is the best-fitting power law and the dashed
  line is the best-fitting de~Vaucouleurs function.  Filled circles
  represent the data from this paper and filled triangles mark points
  calculated from data presented in \citet{lb00}. The dashed
  horizontal line marks a GC surface density of zero and the dotted
  vertical line marks the measured radial extent of the GC system as
  described in Section~\ref{section:radial profile}.}
\label{fig:1023 radial profile}
\end{figure}

\begin{figure}
\plotone{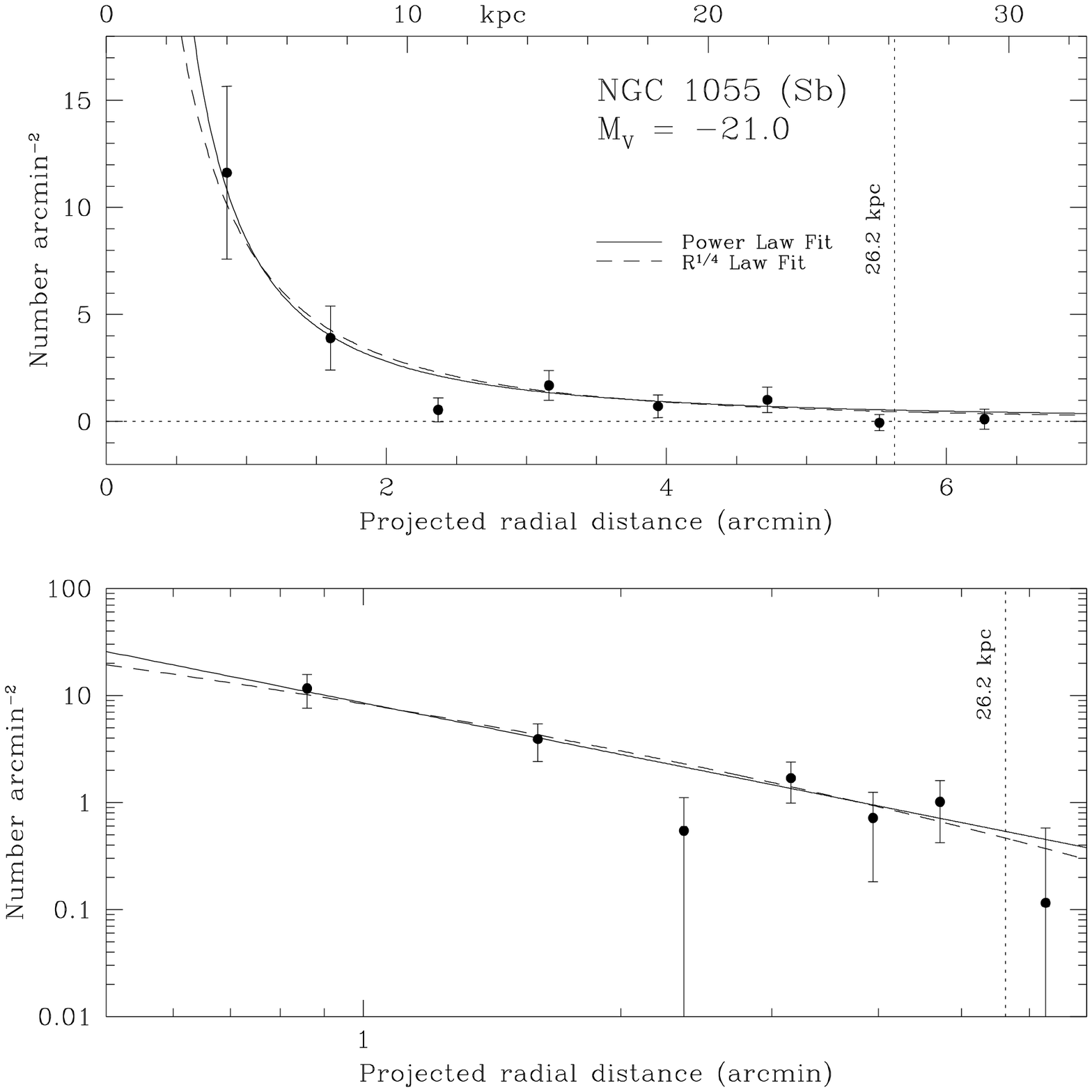}
\caption{ Radial distribution of the GC system of NGC 1055, plotted in
  the same manner as in Figure~\ref{fig:1023 radial profile}.  }
\label{fig:1055 radial profile}
\end{figure}

\begin{figure}
\plotone{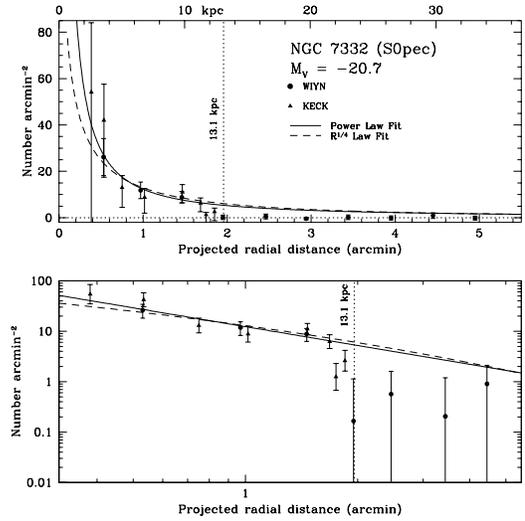}
\caption{ Radial distribution of the GC system of NGC 7332, plotted in
  the same manner as in Figure~\ref{fig:1023 radial profile}. Filled
  circles represent the data points from this paper and filled triangles mark
  points from data presented in \citet{forbes01}.}
\label{fig:7332 radial profile}
\end{figure}

\begin{figure}
\plotone{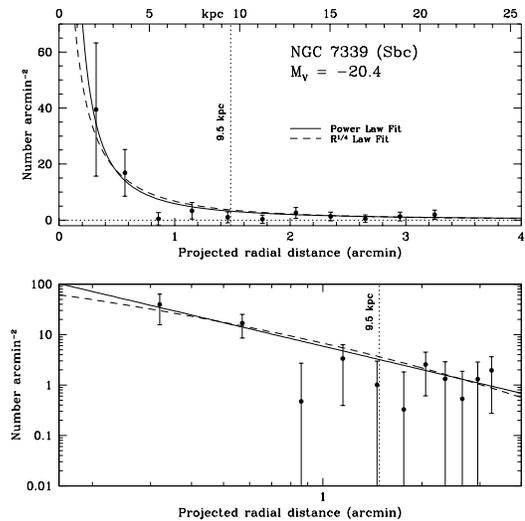}
\caption{ Radial distribution of the GC system of NGC 7339, plotted in
  the same manner as in Figure~\ref{fig:1023 radial profile}.}
\label{fig:7339 radial profile}
\end{figure}

The radial distribution
of the GC system in the final corrected profile falls to zero surface
density (and remains at zero in adjacent bins) within the calculated
errors for each of the target galaxies.  We define the center of this
radial bin as the radial extent of the GC system for that galaxy.  For
NGC~1023, this occurs in the radial bin centered at 6.3\arcm\ and for
NGC~1055 at 5.5\arcm. The GC system radial profiles of NGC~7332 and
NGC~7339 drop to zero surface density within the errors in the
2.0\arcm\ and 1.5\arcm\ annuli, respectively.

It is worthwhile to transform the radial extent of each galaxies' GC
system into physical units of distance.  We calculate this value by
combining the radial extents determined above with the distance moduli
listed in Table \ref{table:galstats}. The errors on the extent values
are calculated for each case by combining the uncertainty in the
galaxy distance modulus with an error equal to one bin width in the GC
system radial profile. For NGC~1023 we find that the GC system extends
out to $20.8 \pm 3.1$ kpc and for NGC~1055 out to $26.2 \pm 6.9$ kpc.
NGC~7332 and NGC~7339 possess less extended GC systems, at $13.1 \pm
3.6$ kpc and $9.5 \pm 2.5$ kpc, respectively.  The measured effective
radii ($R_{\rm eff}$) of the light distributions of our target
galaxies have been tabulated by the Atlas3D Survey (Cappellari et
al.\ 2011). In their data tables, NGC~1023 is listed as having $R_{\rm
  eff}$ $=$ 47.9\arcsec, NGC~1055 has $R_{\rm eff}$ $=$ 67.6\arcsec,
NGC~7332 has $R_{\rm eff}$ $=$ 17.4\arcsec, and NGC~7339 has $R_{\rm
  eff}$ $=$ 22.4\arcsec.  We find, therefore, that the GC systems of
NGC~1023, NGC~1055, NGC~7332, and NGC~7339 extend to approximately
7.8, 4.9, 6.8, and 3.9~$R_{\rm eff}$, respectively.

\begin{figure}
\plotone{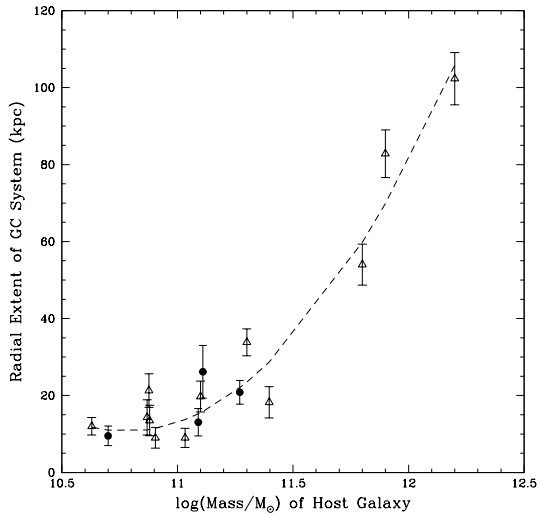}
\caption{Measured radial extent of the GC system in kiloparsecs
  vs.\ the log of the galaxy stellar mass in solar masses for 16 giant
  galaxies included in our wide-field GC system survey so far.  Open
  triangles are points from R10 and H11.  Filled circles are the four
  new points from this paper.  The best-fitting polynomial, which
  agrees within the errors with the curves shown in R07 and R10, is
  plotted as a dashed line.  We note that the fitted function begins
  to curve upward at the low-mass end of the relation.  This may or
  may not be a real effect; more data are needed to quantify the
  behavior of this relation for lower-mass giant galaxies.}
\label{fig:extent mass}
\end{figure}

Our measurements of the GC system radial extent for these four
galaxies fall along the established relationship between the extent of
a GC system and the stellar mass of the host galaxy shown in previous
papers from our wide-field survey (R07, R10).  Fitting a second-order
polynomial to the data including the four new points from this paper
yields a function of the form

\begin{equation}
y = ((47.1\pm5.0)x^2) - ((1020\pm110)x) + (5480\pm620),
\end{equation}
where $x$ is log($M/M_\odot$) and $y$ is the radial extent of the GC
system in kiloparsecs.  The data points and best-fitting function are
plotted in Figure~\ref{fig:extent mass}.

\subsection{Data from Previous Studies}
\label{section:hst data}

For two of the target galaxies, we were able to supplement and
directly compare our study with results from previous work.  In the
case of NGC~1023, S. Larsen and J. Brodie provided us with $V$ and $I$
photometry, positions, and size measurements for 1058 point sources
from their HST WFPC2 study of this galaxy's GC system \citep{lb00}.  To
select GC candidates from the set of point sources, we started by
applying the same color and magnitude cuts as Larsen and Brodie used
in their analysis of NGC~1023's GC population: we selected objects
with $0.75 < (V-I)_0 < 1.40$ and $20 < V < 25$, duplicating their
sample of 221 GC candidates. We decided to implement a more stringent
magnitude cut at the faint end in order to simplify the rest of the
analysis. We removed objects with $V > 24$; at magnitudes brighter
than this threshold, the WFPC2 data are $>$90$-$100\% complete and a
completeness correction is not required. The final list we used
included 151 objects and we assume no magnitude incompleteness in this
list.

Larsen and Brodie assessed the contamination level in their WFPC2 data
by examining a nearby comparison field pointing.  They found five
point sources in the comparison field with $V$ $<$ 24 and $V-I$ colors
like GCs.  Based on this value and the size of the WFPC2 FOV, we
assume the surface density of contaminants in the WFPC2 GC candidate
list is 0.9 $\pm$ 0.4 objects~arcmin$^{-2}$.  Using the same methods as
in Section \ref{section:gclf} we fitted a GCLF to the 151-object WFPC2
sample and found the GCLF coverage to be $79.1\% \pm 3.9\%$.  After
assigning the HST candidates to annular bins and correcting for
missing area, GCLF coverage, and contamination, we added these values
to the WIYN radial profile data to produce the profile shown in Figure
\ref{fig:1023 radial profile} and listed in Table
\ref{table:1023radprof}.  The advantage of combining HST and WIYN data
is apparent in the plot and table: the HST data allow us to probe
deeper into the central regions of the host galaxy, while the WIYN
data enable us to observe the full radial extent of the GC system.

The innermost three points in the WIYN radial profile (between
$\sim$0.75\arcm\ and 2.3\arcm) overlap the radial range of the HST
data, which extends from $\sim$0.4\arcm\ to 2.5\arcm. The WIYN GC
surface density values at 1.5\arcm\ and 2.3\arcm\ closely match the
HST GC surface density values in that radial region, but the innermost
WIYN point at 0.76\arcm\ is low (14.4$\pm$3.0~arcmin$^{-2}$) compared
to the nearby HST point (29.0$\pm$5.6~arcmin$^{-2}$) at a projected
radius of 0.83\arcm.  We thoroughly investigated the reasons for this
discrepancy. A total of 88 HST-identified GC candidates had sky
positions that put them in the same region as our innermost WIYN
radial profile bin.  We marked these positions on our WIYN images,
then individually examined them, looking for matches in the source
lists.  We found that 26 HST-identified GC candidates within the
magnitude detection limits for the WIYN data were located in masked
regions in the WIYN images. An additional seven candidates were in
crowded regions, where sources in the WIYN images were sometimes
blended together; this apparently affected the stellar profiles and
the measured colors enough that these objects were not chosen as GC
candidates.  Based on the WIYN surface density of GC candidates in the
unmasked area, we expected only $\sim$5 objects in the masked regions.
The unfortunate over-density of GC candidates in the masked regions in
this bin explain the low calculated value for the surface density.

For NGC~7332, we were able to compare the results of our radial
profile to those published in \cite{forbes01}.  Forbes et al.\ studied
NGC~7332's GC system with $B$, $V$, and $I$ imaging from the Keck-II
10-m telescope and Low Resolution Imaging Spectrometer (LRIS).  We
have taken surface density values directly from the final GC system
radial profile plotted in their paper and over-plotted them on our
WIYN radial profile in Figure \ref{fig:7332 radial profile}. We also
list their radial profile values along with ours in Table
\ref{table:7332radprof}.  The radial profile from the Forbes et
al.\ study agrees very well with our final WIYN radial profile points,
though the errors on the Forbes et al.\ data are large.

\subsection{Total Number and Specific Frequencies of GCs}
\label{section:spec freq}

\subsubsection{Total Numbers of Globular Clusters}
\label{section:totalN}

A primary goal of our wide-field GC system survey is to determine an
accurate value for the total number of GCs, $N_{GC}$, for each galaxy.
To calculate this number we start by integrating the best-fitting
radial profile function from the innermost bin in the profile to the
outer edge of the bin in which the GC surface density vanishes within
the errors.

In the case of NGC~1055 the best-fitting function (as measured by the
reduced $\chi^2$ value; see Table~\ref{table:fitting}) was a power law
that we integrated from 0.34\arcm\ to 5.9\arcm. The result was 185
GCs. We cannot detect GC candidates inside 0.34\arcm\ (1.6~kpc at the
distance of NGC~1055) because of saturation of the CCD by light from
the galaxy bulge as well as the high noise level in that region.  We
considered three scenarios for the behavior of the radial profile
inside 0.34\arcm: (1) that the power law continues to very small
radius ($r$ very close to 0); (2) that the profile remains flat
(constant surface density) from the innermost populated bin to
$r$$=$0; and (3) that the inner profile is similar to that of the
Milky Way GC system. We found that integrating the power law to very
small radii (close to $r$$=$0) yielded unrealistically large numbers
of GCs in this inner region; it added 85 GCs in a region of radius
only 1.6 kpc, and increased the total number of GCs in the galaxy by
46\%.  Thus we rejected this scenario and averaged the result of the
other two methods. The second possibility was straightforward to
calculate and predicted an additional 17 GCs inside 0.34\arcm, to
yield a total of 202.  To estimate the number of GCs for the third
possibility, we examined the 2010 edition of the McMaster catalog of
Milky Way GCs (Harris 1996) and identified 19 Galactic GCs within
1.6~kpc of the Milky Way center. If one assumes that NGC~1055 has a
similar surface density of GCs within its central 1.6~kpc around the
galaxy center, this yields an extra 28 GCs in the central region and a
final total of 213 GCs for the system. The average result for the two
options (flat or Milky-Way-like surface density inside 0.34\arcm)
yields 207.5 GCs. We computed an error on $N_{GC}$ by adding in
quadrature the errors due to fractional coverage of the GCLF, Poisson
errors on the number of GCs and contaminating objects, and uncertainty
in the number of GCs in the central region of the galaxy. Our final
value (rounded to significant digits) for the total number of GCs in
NGC~1055 is $N_{GC}$ $=$ $210 \pm 40$.

We executed the analogous steps for NGC~1023 and integrated the
best-fitting de~Vaucouleurs law function from 0.27\arcm\ to
6.67\arcm\ to derive the number of GCs (470) in that region.  Because
we had made use of the HST data from \citet{lb00}, we were able to
start the integration at a smaller radius (0.27\arcm, or
$\sim$0.90~kpc at the distance of NGC~1023) than if we had only had
the WIYN imaging.  Inside 0.27\arcm, where we had no observations, we
averaged the value obtained from continuing the de~Vaucouleurs
integration to the center (an additional 26 GCs), and the value
calculated from assuming a constant surface density to $r$$=$0 (16
GCs).  It seemed appropriate and valid to average the result from
these two methods because the de~Vaucouleurs law does not become
unphysical at small radii, and the two options yielded similar results
in any case.  The outer integration edge was set at the point where
the radial profile drops to levels consistent with zero.  For
NGC~1023, we calculate a final total $N_{GC}$ of $490 \pm 30$.

A similar analysis was performed on the radial profiles of NGC~7332
and NGC~7339.  In both cases a power law profile yielded the best fit
to the radial profile data; these functions were integrated from
0.2\arcm\ to 2.2\arcm\ for NGC~7332, and 0.1\arcm\ out to
1.6\arcm\ for NGC~7339 to calculate the number of GCs in those radial
ranges.  The result was 155 for NGC~7332 and 71 for NGC~7339.  To
determine how many GCs were likely to be inside the innermost radial
bin, we integrated the power law functions to very small radii ($r$
close to 0) and found that they gave reasonable results this time (for
example, integrating the de~Vaucouleurs law fit all the way to the
galaxy center gave results very similar to the power law integration
for both galaxies). For NGC~7332, the flat profile yielded 11
additional GCs in the central region, while the power-law profile gave
a value of 27 GCs over the same region.  For NGC~7339, the flat
profile predicted three more GCs in the inner region while the power
law profile predicted five more GCs.  The final estimates of $N_{GC}$
$=$ $175 \pm 15$ for NGC~7332 and $75 \pm 10$ for NGC 7339 are the
result of taking the mean of the values obtained when assuming a flat
profile inside the innermost bin and a continuous power law into the
center of the target galaxy (very near $r$ $=$ 0).

\subsubsection{Luminosity- and Mass-Normalized Specific Frequencies}
\label{section:sn_and_t}

A useful value for characterizing the GC system of a galaxy is the
specific frequency, $S_N$, or the total number of GCs normalized by
the host galaxy luminosity.  Specific frequency was first defined by
\citet{hvdb81} as a tool for investigating whether the ability of a
galaxy to form GCs depends on the galaxy's luminosity or on one or
more other factors.  It is defined by the equation \begin{equation}S_N
  \equiv N_{GC}~10^{+0.4(M_V +15)}\end{equation} \citep{hvdb81}, where
$N_{GC}$ is the total number of GCs and $M_V$ is the galaxy's absolute
$V$ magnitude. Combining the final determinations of $N_{GC}$ with the
$M_V$ values listed in Table \ref{table:galstats} yields $S_N$ $=$
$1.7 \pm 0.3$ for NGC~1023, $S_N$ $=$ $0.9 \pm 0.2$ for NGC~1055,
$S_N$ $=$ $0.9 \pm 0.3$ for NGC~7332, and $S_N$ $=$ $0.5 \pm 0.2$ for
NGC~7339.

The quantity $S_N$ has a stellar population dependence --- i.e.,
variations in the $V$-band stellar mass-to-light ratios for galaxies
of different morphological types will contribute to differences in
$S_N$.  This in turn complicates the comparison of 
specific frequency values among galaxies of various types. To address
this issue, \citet{za93} formulated the $T$ parameter, which is the
number of GCs normalized by the stellar mass of the parent galaxy and
is defined by
\begin{equation}T \equiv
\frac{N_{GC}}{M_G/10^9M_{\odot}}\end{equation} \citep{za93}, where
$N_{GC}$ is again the total number of GCs and $M_G$ is the stellar
mass of the galaxy.  Table \ref{table:galstats} lists the calculated
$M_G$ for each galaxy based on the given $M_V$ and an assumed $V$-band
mass-to-light ratio value that varies with morphological type as
defined in \citet{za93}.  The final $T$ values for the four galaxy
targets are $2.7 \pm 0.4$, for NGC~1023, $1.6 \pm 0.5$ for NGC 1055,
$1.4 \pm 0.4$ for NGC~7332, and $1.5 \pm 0.5$ for NGC~7339.  Errors
for $S_N$ and $T$ were computed by propagating the error on $N_{GC}$
and including the errors on the total galaxy magnitudes. The final
results for $N_{GC}$, $S_N$, and $T$ for these four galaxies are
included in the first four entries in Table~\ref{table:results} in
columns 4, 5, and 6, respectively.  The other galaxies included in
Table \ref{table:results} are introduced and discussed in Section
\ref{section:comparison} and the quantity in column (7), $T_{\rm blue}$,
is discussed in Section \ref{section:tblue}.

\subsubsection{Previous Estimates of $N_{GC}$ and $S_N$ for the Target Galaxies}
\label{section:previous}

Only one of the four target galaxies, NGC~7332, has estimates of GC
specific frequency already published in the literature.  We searched
the literature and found no other published articles on the GC
populations of NGC~1055 or NGC~7339.  The \citet{lb00} paper on the GC
system of NGC~1023 did not include estimates of the total population
or radial extent of the full GC system.  Their study was constrained
by the small field-of-view of HST/WFPC2 and the relative proximity of
NGC~1023, which meant that their spatial coverage was very limited.

In their Keck-II imaging study, \citet{forbes01} estimated that the
total number of GCs in NGC~7332 is $190 \pm 30$; this is consistent
with our final derived value of $175 \pm 15$.  \citet{forbes01}
calculated a specific frequency $S_N$ $=$ $2.0 \pm 0.3$, whereas we
calculate $S_N$ $=$ $0.9 \pm 0.3$.  Since our numbers for $N_{GC}$ are
consistent, this difference in $S_N$ is due to Forbes \et using a much
smaller distance to the galaxy.  Their assumed distance is 15.3~Mpc
and is derived by combining the redshift and assuming $H_0$ $=$ 75
\kms~$Mpc^{-1}$ with an additional Virgocentric infall correction.
Combining the distance derived from surface brightness fluctuations
(23~Mpc; Tonry et al.\ 2001) with 
their $N_{GC}$ number yields $S_N$ $=$ $1.0 \pm 0.3$, which is
consistent with our result.

\subsubsection{Comparisons of Total Numbers and Specific Frequencies to Those of Other Giant Galaxies}  
\label{section:comparison}

Our GC system survey is ongoing and we add new observations of giant
galaxies and their GC populations each observing season.  Subsequent
papers are planned that will include results for several more
early-type galaxies (Hargis \& Rhode 2012a, in preparation) and
present a thorough multivariate analysis of how GC system properties
vary with host galaxy luminosity, mass, morphology, and environment
(Hargis \& Rhode 2012b, in preparation).  For now, we can examine how
the four spiral and S0 galaxies analyzed here compare to the other
spiral and E/S0 galaxies for which secure global GC system properties
have been measured.

In Table \ref{table:results} we have compiled the measured global
properties of the GC systems of a sample of spiral, S0, and elliptical
galaxies, including morphological type, $M_V$, $N_{GC}$, $S_N$, $T$,
and $T_{\rm blue}$ (see Section \ref{section:tblue}).  The data for
thirteen of the galaxies are taken from our ongoing GC system survey
(RZ01, RZ03, RZ04, R07, R10, H11, Hargis \& Rhode 2012a, in
preparation).  We supplement the survey data with data from other
wide-field, multi-color imaging studies of galaxy GC system properties
in the literature (NGC~1052 from \citealp{forbes01}; NGC~4374 from
\citealp{gr04}; NGC~5128 from \citealp{harris06}, with the blue GC
fraction adopted from \citealp{harris04}).  We also include values for
the GC systems of the Milky Way \citep{az98} and M31 (\citealp{az98},
\citealp{barmby00}, \citealp{perrett02}).\footnote{We note that
  surveys of star clusters in M31 are ongoing and recent studies have
  indicated that the number of massive ``globular-like'' clusters may
  be as large as $\sim$650 (Revised Bologna Catalog, Galleti et
  al.\ 2004).  On the other hand, Fan, de~Grijs, \& Zhou (2010) select
  a sample of objects from the Revised Bologna Catalog and find 445
  confirmed GCs in M31.  In any case, increasing $N_{GC}$ for M31 from
  450$\pm$100 to 650$\pm$150, and thereby increasing the GC specific
  frequencies for that galaxy, does not change the results of the
  comparison presented here and in Section~\ref{section:tblue}.}  This
gives us a total of 18 galaxies (hereafter referred to as the $N$$=$18
sample, which is made up of half spiral galaxies and half E/S0
galaxies) with well-measured $S_N$ and $T$ to compare to the values
measured for the four galaxies in the current paper.

The weighted mean $S_N$ and $T$ values for the nine spiral galaxies in
our $N$$=$18 comparison sample are 
%
$0.5 \pm 0.1$ and $1.1 \pm 0.1$, respectively. The $T$ values for
NGC~1055 and NGC~7339, and the $S_N$ for NGC~7339 are consistent with
the spirals in the $N$$=$18 sample, while the $S_N$ for NGC~1055 falls
slightly above, but within $2\sigma$ of, the weighted mean.  Including
the results for NGC~1055 and NGC~7339 (for a total of 11 spiral
galaxies) in the weighted mean gives us $S_N = 0.6 \pm 0.1$ and $T =
1.2 \pm 0.1$.

The nine E/S0 galaxies in the $N$$=$18 sample have weighted mean
values of $S_N = 1.6 \pm 0.1$ and $T = 2.4 \pm 0.2$.
The $S_N$ and $T$ for NGC~1023 are typical of the larger sample,
whereas NGC~7332 falls below the weighted mean in both categories, but
within the large range of measured $T$ values seen in the comparison
galaxies.  For the E/S0 galaxies in the $N$$=$18 sample, the GC
specific frequencies are in the range $S_N$ of $0.8-3.6$ and
$T$ of $1.2-4.8$.  Adding our results for NGC~1023 and NGC~7332 to
the weighted mean (for a total of 11 E/S0 galaxies) results in a new
weighted mean $S_N = 1.6 \pm 0.1$ and $T = 2.3 \pm 0.2$.

As we add more galaxies to our sample, we will be able to divide the
galaxies more finely according to their properties (e.g., specific
morphological type and environment) and look for trends of GC system
properties with host galaxy properties.  For now, given the above, we
can say with confidence that in terms of $S_N$ and $T$, NGC~1023 and
NGC~7332 possess values consistent with other early-type galaxies with
carefully-measured global GC specific frequencies.  NGC~1055 and
NGC~7339 likewise fall within the normal range of GC specific
frequencies observed for spiral galaxies.

\subsection{The GC System Color Distributions and the
  \\Specific Frequency of Blue GCs}
\label{section:tblue}

Quantifying both the GC system color distribution and the blue GC
specific frequency (the number of blue, metal-poor GCs normalized by
the galaxy mass) of the target galaxies are two of the goals of the
overall GC system survey.  In old stellar populations (with ages
greater than $\sim$2~Gyr), differences in broadband optical color are
primarily due to differences in metallicity (e.g., \citealp{brodie06}
and many references therein). Metal-poor GCs would therefore have
bluer optical colors than metal-rich GCs.  The GC systems of many
giant galaxies, including the Milky Way \citep{zinn85}, M31
(\citealp{barmby00}, \citealp{perrett02}), and many ellipticals
(\citealp{kw01}, \citealp{kz07}, \citealp{strader07}) have been shown
to include two (or sometimes more) populations of GCs with different
metallicities, likely formed in different episodes of star
formation. Various galaxy formation models (e.g., \citealp{az92},
\citealp{fbg97}, \citealp{cote98}, \citealp{beasley02},
\citealp{mg10}) make predictions for the relative numbers of blue and
red GCs and how those numbers might vary in different types of
galaxies and at different radii within galaxies.  One interesting
piece of analysis that we can do, therefore, is to examine the numbers
of blue, metal-poor GCs in our target galaxies and compare them to the
values for other galaxies. In general, galaxies with higher stellar
mass tend to have higher specific frequencies of blue, metal-poor GCs
(e.g., R05, R07, Peng et al.\ 2008, Spitler et al.\ 2008).  Moreover,
the high blue GC specific frequencies of the most massive giant
elliptical galaxies suggest that they are unlikely to have formed from
the straightforward merger of typical spiral galaxies and their GC
populations (e.g., \citealp{harris03}, R05, \citealp{brodie06}).

For each of the four target galaxies, we set out to examine the
morphology of the GC color distribution, determine the fraction of
blue and red GCs in the galaxy, and calculate $T_{\rm blue}$, which is
defined in the same way as $T$ was defined in
Section~\ref{section:sn_and_t} except that $N_{\rm GC}$ this time
includes only the blue portion of the GC population. To ensure no bias
(i.e., toward red or blue objects) we begin by constructing a list
that includes only those GC candidates that lie within the 90\%
complete magnitude ranges for all three filters (see RZ01 for more
details); we refer to this as the 90\% complete sample.

For NGC~1023, the 90\% complete sample contains 119 objects.  We used
this sample as input to the KMM mixture modeling code (Ashman \et
1994) to test for bimodality in the $B-R$ color distribution.  Based
on the GC candidates' $B-R$ color, the KMM code found that unimodal
fit was rejected at greater than 95\% confidence.  The resulting
bimodal distribution puts 68 out of 119 objects (57\%) in the blue
(lower metallicity) peak.  Taking the fraction of blue GCs in the
total sample from this estimate and combining it with $N_{GC}$ for
NGC~1023 and the stellar mass value from Table~\ref{table:galstats},
we find a mass-normalized number of metal-poor GCs of $T_{\rm blue} =
1.5 \pm 0.2$.  In typical giant galaxies with double-peaked GC color
distributions, the dividing line between the blue- and red-dominated
regions occurs at $B-R \sim 1.23$ (R01, R04).  We find reasonable
agreement with that expectation here, as $B-R \sim 1.26$ marks the
overlap of the two populations in NGC~1023.

\begin{figure}
\plotone{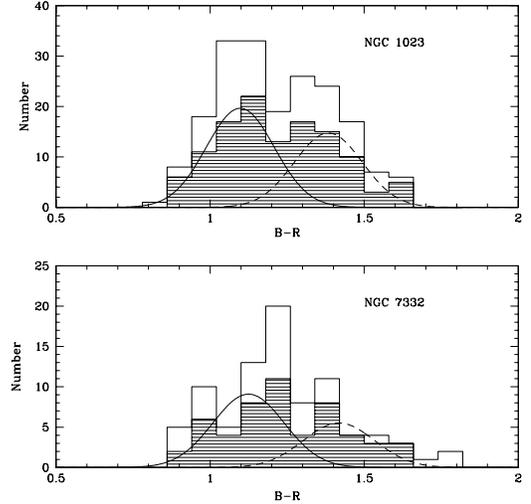}
\caption{Color distributions for GC candidates in NGC~1023 and
  NGC~7332.  The top panel illustrates the $B-R$ color distribution
  for GC candidates NGC~1023.  The histogram of the full 192 GC
  candidate list is marked by the solid line, and the 90\% complete
  sample of 119 objects is indicated by the shaded region.  The curves
  are the output of the KMM fitting routine taking the 90\% complete
  sample as input, with the solid curve marking the blue, metal-poor
  peak while the dashed curve marks the red, metal-rich peak of GC
  candidates.  The bottom panel follows a similar convention, showing
  the color distribution of the GC candidates in NGC~7332, with 90 GC
  candidates in the full list, and 53 in the 90\% complete sample.}
\label{fig:color_dist}
\end{figure}

A similar analysis was applied to the GC population of NGC~7332.  In
this case, the 90\% complete sample contained 53 objects, just above
the 50-object lower limit for the KMM algorithm.  Applying KMM to this
sample resulted in a weak rejection of a unimodal distribution at an
81.2\% confidence level.  For the bimodal case, KMM places 33 out of
53 objects (62\%) in the blue portion of the distribution.  The line
marking the overlap between the two regions occurs at $B-R \sim 1.30$.
Applying the 62\% fraction to our full sample gives us $T_{\rm blue} =
0.8 \pm 0.3$.  If we instead simply split the 90\% sample into blue
and red GC candidates at the typical dividing line for massive
elliptical galaxies of $B-R$ $=$ 1.23 (RZ01, RZ04), the proportions of
blue and red GCs in NGC~7332 are 51\% and 49\%, respectively. The
$T_{\rm blue}$ value then becomes 0.7$\pm$0.2.
Figure~\ref{fig:color_dist} illustrates the color distributions for GC
candidates in NGC~1023 (top) and NGC~7332 (bottom).  The unshaded
region marks our full sample, while the shaded region contains the
90\% complete sample.  The solid curve marks the bluer, more
metal-poor population, while the dashed curve marks the redder, more
metal-rich population.

For NGC~1055, our 90\% complete sample contained only 13 objects.
Because this sample is so small, we were unable to use the KMM
algorithm to investigate whether two populations might be present.
Instead, we can look at what happens when we divide the population
along the previously referenced $B-R$ $\sim$ 1.23 criterion.  Six of
the 13 (46\%) objects in the 90\% complete color sample for NGC~1055
are bluer than $B-R$ $=$ 1.23.  If we assume that the results from
this small sample are representative of the entire GC system, we find
that $T_{\rm blue}$ for NGC~1055 is 0.75$\pm$0.21.  If we assume
instead that NGC~1055 has a blue GC fraction more like that of the
Milky Way or M31 ($\sim$70\%; \citealp{harris96}, \citealp{barmby00},
\citealp{perrett02}), the $T_{\rm blue}$ value becomes 1.14$\pm$0.32.
The average of these two values (rounded to significant digits) is
0.9$\pm$0.3, which we will take as our final estimate of $T_{\rm
  blue}$ for NGC~1055.

The 90\% complete sample for NGC~7339 also contained too few objects
to use with KMM.  We find that for NGC~7339, 21 out of 39 objects
(54\%) fall on the blue side of $B-R \sim 1.23$, giving a $T_{\rm
  blue} = 0.8 \pm 0.3$.  If we assume a 70\% blue fraction like the
Milky Way or M31, $T_{\rm blue}$ becomes 1.1$\pm$0.4.  The average of
these two values is 0.9$\pm$0.3.

As in Section \ref{section:comparison} we can compare and integrate
our results for $T_{\rm blue}$ with well-determined values for other
galaxies taken from our ongoing survey and the literature (see Table
\ref{table:results}).  The weighted mean $T_{\rm blue}$ for the nine
spiral galaxies in the $N$$=$18 sample is $0.8 \pm 0.1$, which is
consistent with our measurements for the spiral galaxies NGC~1055 and
NGC~7339.  
Including the $T_{\rm blue}$ measurements for NGC~1055 and
NGC~7339 does not change the weighted mean: it stays at 
$T_{\rm blue}$ $= 0.8 \pm 0.1$. 
The nine E/S0 galaxies in the $N$$=$18 sample have a weighted mean
$T_{\rm blue}$ $= 1.4 \pm 0.1$.  Adding in our new $T_{\rm blue}$
estimates for NGC~1023 and NGC~7332 again does not change the weighted
mean within the errors: the mean $T_{\rm blue}$ for all eleven E/S0
galaxies is $1.3 \pm 0.1$.  The {\it range} of $T_{\rm blue}$ for
early-type galaxies in the sample is fairly large, 0.7$-$2.6, with
massive cluster ellipticals having $T_{\rm blue}$ greater than
$\sim$2.

The general trend identified in R05 --- i.e., a rough increase in the
number of blue, metal-poor GCs per galaxy mass with increasing galaxy
stellar mass --- still holds with the addition of the four galaxies
analyzed here.  This type of trend is consistent with the biased,
hierarchical scenario of GC and galaxy formation outlined in
\citet{santos03} and R05.  In this type of picture, today's massive
galaxies located in dense environments have comparatively high numbers
of blue GCs for their stellar mass.  This is because a larger
proportion of their mass was in place at high redshift (massive
galaxies began the collapse and assembly process earlier than
lower-mass galaxies in low-density environments) and thus could
participate in star cluster formation when the first generation of GCs
was forming in the Universe.  Detailed discussions of $T_{\rm blue}$ and
possible scenarios that would give rise to the differences in
$T_{\rm blue}$ values that we see are given in R05, R07 and R10; suffice
it to say here that the spiral and S0 galaxies in the current paper
appear to have fairly typical $T_{\rm blue}$ for their masses and
morphological types.

\begin{figure}
\plotone{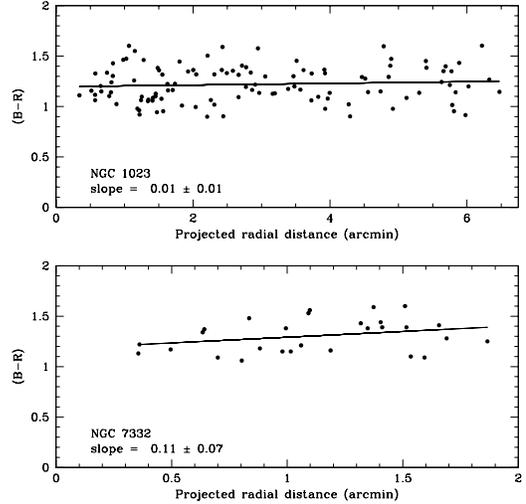}
\caption{ Color gradient of the GC systems of NGC~1023 (top) and
  NGC~7332 (bottom).  The B-R color vs. galactocentric projected
  radius of the GC candidates in the 90\% complete sample
  (Section~\ref{section:tblue}), plotted out to the radial extent of
  the GC system for the galaxy (Section~\ref{section:radial profile}).
  The best-fitting least-squares line is plotted and the slope and
  associated error is specified on the plot for each system. No
  significant color gradient is found for either GC system.}
\label{fig:color_gradient} 
\end{figure}

Lastly, we note that as part of the analysis of the GC color
distributions, we investigated whether any color gradients are present
in the galaxies' GC systems.  If the red, metal-rich GC population is
less spatially extended compared to the blue, metal-poor GC population
(in other words, the ratio of red to blue GCs decreases with
increasing radial distance from the galaxy center), one might expect
to see a measurable color gradient in the GC system as a whole.  The
presence of a color gradient could imply that the gas from which the
red GCs formed was subject to additional dissipational collapse
compared to the gas from which the blue GCs formed (e.g., AZ92, Forbes
et al.\ 1997, Beasley et al.\ 2002) and/or that extra blue GCs are
present in the galaxies' halo that were accreted from metal-poor dwarf
galaxies (e.g., C\^ot\'e et al. 1998).
In any case, only NGC~1023 and NGC~7332 had enough objects in the 90\%
sample to do this type of analysis. We analyzed the $B-R$ color of the
GC candidates as a function of projected radius and found no
significant evidence for a radial color gradient in either galaxy's GC
system.  Figure~\ref{fig:color_gradient} shows the GC candidate colors
versus projected radius for the 90\% sample of NGC~1023 (top) and
NGC~7332 (bottom) out to the radial extent of the GC system as
determined in Section~\ref{section:radial profile}.  The best fit
least-squares line for each sample is plotted, with the slope
determined by this fit indicated at the bottom of each plot.

\subsection{GC Candidates in the Dwarf Galaxy NGC~1023A}
\label{section:dwarf}

As explained in Section~\ref{section:final selection}, during the GC
candidate selection process for NGC~1023 we masked out a circular
region with radius 0.6\arcm\ around the nearby dwarf satellite galaxy
NGC~1023A.  Eight GC candidates (point sources with magnitudes and
colors like GCs) were located within the masked area.  To determine
what fraction of the eight objects are likely to be truly associated
with the dwarf galaxy NGC~1023A we must first consider possible
sources of contamination.  We estimated the surface density of
contaminants (i.e., foreground stars and background galaxies) in the
WIYN image based on the asymptotic behavior of the profile of
NGC~1023's GC system and derived a contamination level of $0.74 \pm
0.20$ objects~arcmin$^{-2}$ (Section \ref{section:asymptotic}).
Combining this surface density with the area of the circular masked
region around the dwarf galaxy implies that 0.80$\pm$0.21
contaminating objects should lie within that region.  We must also
consider the GC system of NGC~1023 as a source of ``contaminants''
within the area around NGC~1023A.  Accordingly, we used the
de~Vaucouleurs fit to the radial profile of the GC system of NGC~1023
(Table \ref{table:fitting}) to calculate a surface density of GCs for
each pixel in the masked region of NGC~1023A.  Multiplying this value
by the area of each pixel and summing the values of all the pixels in
the masked region together yields the total contribution from the GC
system of NGC~1023 in the masked region. Because we are concerned
about the number of {\it detected} objects, we undo the GCLF
completeness scaling factor that was used to construct the final
radial profile of NGC~1023 (Section \ref{section:gclf}), yielding a
value of $3.1 \pm 1.3$ detectable objects from the GC system of
NGC~1023 coincident with the spatial position of the dwarf galaxy
NGC~1023A.  Considering both the contamination from foreground and
background objects and from the NGC~1023 GC system, we expect a total
contamination value of $3.9 \pm 1.4$ objects.

\begin{figure}
\plotone{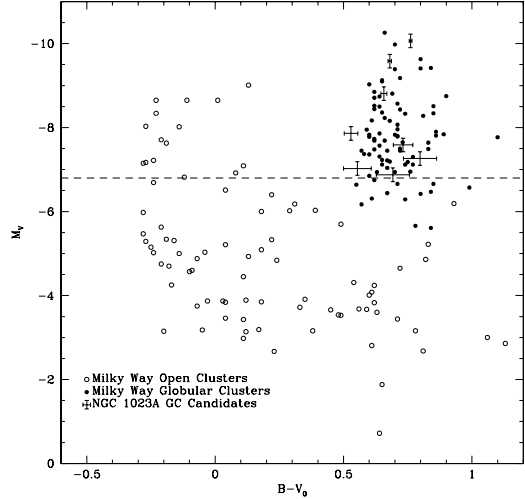}
\caption{Integrated absolute $V$-band magnitude, $M_V$, versus $B-V_0$
  color, for Milky Way open clusters (open circles), Milky Way
  globular clusters (filled circles), and the GC candidates in
  NGC~1023A (marked with error bars).  Open cluster values were
  obtained from \citet{lps02} and globular cluster values come from
  \citet{harris96}.  Error bars for NGC~1023A GC candidates include
  photometric and distance modulus errors.  The dashed line marks the
  faint magnitude threshold for GC selection in NGC~1023, which we set
  at $M_V = -6.8$ ($V$$=$23.5; see Section \ref{section:final
    selection}).  Above the faint magnitude threshold there is a clear
  separation in $B-V$ color between open clusters and globular
  clusters.}
\label{fig:color_mag_n1023a} 
\end{figure}

One more issue needs to be addressed before we can estimate the number
of GCs in NGC~1023A.  This galaxy is a dwarf irregular (dIrr), so it
has ongoing star formation.  In fact, \citet{lb02} spectroscopically
confirmed the presence of two young ($\sim$125$-$500~Myr) blue
clusters in the central region of NGC~1023A, leading them to speculate
that a recent ($< 500$ Myr) close encounter with NGC~1023 may have led
to an enhanced period of star formation in the dwarf galaxy.  Young
clusters like these may represent another source of contamination in
our sample of GC candidates in NGC~1023A -- young, blue open clusters
might have similar magnitudes and colors to GCs.  To investigate this
possibility, we have plotted in Figure~\ref{fig:color_mag_n1023a} the
integrated $M_V$ vs. $B-V_0$ for Milky Way open clusters
\citep{lps02}, Milky Way globular clusters \citep{harris96}, and the
eight GC candidates for NGC~1023A.  $M_V$ for the eight GC candidates
in our sample is calculated by assuming the same distance to NGC~1023A
as to NGC~1023 (Table~\ref{table:galstats}) and errors are propagated
from the distance modulus and photometric errors.
 The dashed line marks the faint-end cut in our sample at $M_V > -6.8$
 (Section \ref{section:final selection}).  Below this line there is an
 area where the lowest-luminosity globular clusters and the most
 luminous, reddest open clusters do overlap.  Within our selection
 region however, there is a clear separation between the Milky Way
 open clusters and globular clusters. Thus we do not consider open
 clusters to be a source of contamination in our GC candidate sample
 for NGC~1023A. The number of GC candidates in NGC~1023A corrected for
contamination is therefore 4.1$\pm$1.3.

Lastly, to produce a final estimated number of GCs in NGC~1023A, we
need to correct for the magnitude incompleteness (GCLF coverage) of
our data.  \citet{gphg09} studied the GC systems of 30 dIrr galaxies
and fitted the combined GCLF of the systems with a Gaussian function
with a peak at $M_V = -7.56 \pm 0.02$ and a dispersion of $1.23 \pm
0.03$.  Our number of detected GC candidates is far too small to
attempt to directly fit a GCLF to our data.  However, we note that the
GCLF peak in dIrr galaxies is brighter than the GCLF peak for giant
galaxies ($M_V = -7.33$; \citealp{az98}).  Given that we found our
coverage of the GCLF in the GC system of NGC~1023 to be 57\% (Section
\ref{section:gclf}), we assume at a \textit{minimum} the same level of
coverage of the GCLF of NGC~1023A.  In an alternate scenario, we
assume we have detected every GC in NGC~1023A.  The 57\%- and
100\%-complete scenarios mark the range of possible values for our
coverage of the GCLF.  We average the result from these two
possibilities and calculate that NGC~1023A has $N_{GC} = 5.7 \pm 2.5$.
The error on this value was determined by adding in quadrature the
Poisson errors on the number of detected objects, the errors on the
contamination estimates, and the uncertainty in the assumed GCLF
coverage scenarios. \citet{gh76} published two estimates of total $V$
magnitude of NGC~1023A based on photoelectric photometry within two
different apertures: $V_T^0$ $=$ 15.27 $\pm$ 0.03 for a
28$\arcs$-diameter aperture and $V_T^0$ $=$ 14.22 $\pm$0.05 for a
45$\arcs$-diameter aperture.  Using the distance modulus for NGC~1023
(Table \ref{table:galstats}) and taking the difference in values as
the error, we adopt a value of $M_V^T = -15.55 \pm 0.53$.
Calculating the luminosity-weighted specific frequency (Section
\ref{section:sn_and_t}) we find that $S_N = 3.5 \pm 2.5$. The error on
$S_N$ was determined by propagating the error on $N_{GC}$ and
including the the uncertainty in the galaxy magnitude $M_V^T$.
\citet{ggph08} found a very broad range of $S_N$ for 13 dIrr galaxies,
spanning values from 0.3 to 11.  The position of NGC~1023A toward the
low end of this $S_N$ range indicates that the integrated luminosity
is likely dominated by the young stellar population found by
\citet{lb02}.  As this population fades, the $S_N$ value for NGC~1023A
will increase.

\section{Summary}
\label{section:summary}

We have obtained three-filter imaging with the 4096 x 4096-pixel
Minimosaic CCD imager on the WIYN 3.5-m telescope in order to study
the GC populations of four S0 and spiral galaxies out to large
galactocentric radii ($\sim$10$-$30~kpc). When possible, we combined
the WIYN imaging data with archival and published HST/WFPC2 and
Keck-II/LRIS data to help quantify the global properties of the
galaxies' GC populations.

We use the WIYN imaging and supplementary data to carefully construct
the final radial distribution (surface density of GCs vs.\ projected
radius) of each galaxy's GC system.  We then integrate the
distributions to derive the total number of GCs ($N_{\rm GC}$).  The
derived total numbers and specific frequencies for the four target
galaxies, as well as values for other galaxies from our ongoing survey
and from other published sources, are given in Table
\ref{table:results}.  The number of GCs ranges from 75$\pm$10 for the
least-massive galaxy of the set, the Sbc galaxy NGC~7339, to
490$\pm$30 for the massive S0 galaxy NGC~1023.  The GC specific
frequencies for these spiral and S0 galaxies are comparable to the
mean values of other well-studied spiral and lenticular galaxies from
our survey and from the literature.

We find that the $B-R$ color distribution of GCs in NGC~1023 is
significantly bimodal ($>$95\% confidence), whereas the color
distribution of NGC~7332's GC population is bimodal at only the
$\sim$81\% confidence level. The blue, metal-poor GCs in these
galaxies make up
$\sim$60\% of of the total population and the blue GC specific
frequencies are consistent with the range of values expected for the
galaxies' morphological types.  We look for the presence of color
gradients due to the changing ratio of blue to red GCs in the two
galaxies with sufficient numbers of GC candidates to do this type of
analysis and find no significant gradients in the GC systems.

We have identified a subsample of eight GC candidates that coincide
with the location of the dIrr galaxy NGC~1023A, which lies
$\sim$2.5\arcm\ ($\sim$8~kpc) in projected distance from NGC~1023.
After accounting for possible contamination from the GC population of
NGC~1023 and foreground and background sources, we estimate the dwarf
galaxy has a total population of GCs equaling $N_{GC} = 5.7 \pm 2.5$
and a GC specific frequency $S_N = 3.5 \pm 2.5$.

Both data acquisition and analysis for the wide-field GC system survey
continue and we add new galaxies to the data set each semester.  The
next two papers from the survey will describe the results for several
elliptical and lenticular galaxies located both in the field and in
galaxy clusters (Hargis \& Rhode 2012a, in preparation) and present a
multi-variate analysis of the current survey sample as a whole (Hargis
\& Rhode 2012b, in preparation).
%

\acknowledgments We thank S\o ren Larsen and Jean Brodie for providing
us with the source list from their 2000 HST WFPC2 study of
NGC~1023's GC system.  The research described in this paper was
supported by an NSF Faculty Early Career Development (CAREER) award
(AST-0847109) to KLR and also by a graduate fellowship from the
Indiana Space Grant Consortium to JLD.  We are grateful to the staff
at the WIYN Observatory and Kitt Peak National Observatory for their
assistance during our observing runs. We thank an anonymous referee
for providing a detailed report with many useful suggestions for
improvements to the paper.  This research has made use of the
NASA/IPAC Extragalactic Database (NED) which is operated by the Jet
Propulsion Laboratory, California Institute of Technology, under
contract with the National Aeronautics and Space Administration.



\begin{deluxetable}{lcrcccc}
\tablecaption{Basic Properties of the Target Galaxies}
\tablehead{\colhead{Galaxy} &\colhead{Type} &\colhead{$m-M$} &\colhead{Distance (Mpc)} &\colhead{$M_V^T$} &\colhead{$\log{M/M_\odot}$} &\colhead{Environment} } 
\startdata
NGC~1023 & S0    & $30.29 \pm 0.16$ & 11.4 & $-$21.2 & 11.3 & Group\\
NGC~1055 & Sb    & $31.06 \pm 0.47$ & 16.3 & $-$21.0 & 11.1 & Group\\
NGC~7332 & S0pec & $31.81 \pm 0.20$ & 23.0 & $-$20.7 & 11.1 & Isolated Pair\\
NGC~7339 & Sbc   & $31.75 \pm 0.37$ & 22.4 & $-$20.4 & 10.7 & Isolated Pair\\
\enddata

\tablecomments{ Morphological types are from RC3
  \citep{devauc91}.  Distance to NGC~1055 is from Willick et
  al. (1997, Tully-Fisher relation). Distance to NGC~7339 is from
  Tully et al. (2009, Tully-Fisher relation).  Distances to NGC~1023
  and NGC~7332 are from Tonry et al. (2001, surface brightness
  fluctuations).  Total absolute magnitudes are from combining $V_T^0$
  from RC3 with $m-M$.  Values of $\log{M/M_\odot}$ were calculated
  using $M_V^T$ and an assumed morphologically-dependent mass-to-light
  ratio, $(M/L)_V$, of 7.6 for NGC~1023 and NGC~7332, 6.1 for
  NGC~1055, and 5.0 for NGC~7339 \citep{za93}.  Environment
  descriptions are from \citet{devauc76}.}
\protect\label{table:galstats}
\end{deluxetable}

\begin{deluxetable}{llrrr}
\tablecaption{WIYN Minimosaic Observations of the Target Galaxies}
\tablewidth{0pt}
\tablehead{\colhead{Galaxy} &\colhead{Date}&\multicolumn{3}{c}{Exposure Times (s)} \\
\colhead{} & \colhead{} & \colhead{$B$} & \colhead{$V$} & \colhead{$R$}}
\startdata
NGC~7332/NGC~7339 & 2008 Sep & 4 x 2100 & 3 x 2000 & 6 x 1800 \\
NGC~1055 & 2008 Sep & 4 x 2100 & 3 x 2000 & 3 x 1800 \\
NGC~1023 & 2009 Sep & 4 x 2100 & 4 x 2000 & 5 x 1800\\
\enddata
\protect\label{table:observations}
\end{deluxetable}

\begin{deluxetable}{lcccccccc}
\tablecaption{Bright GC Candidates in NGC~1023}
\tablehead{
\colhead{Sequence \#}& \colhead{RA (2000)}& \colhead{Dec (2000)}& \colhead{$V$}& \colhead{$\sigma_{V}$}& \colhead{$B-V$}& \colhead{$\sigma_{B-V}$} & \colhead{$V-R$}& \colhead{$\sigma_{V-R}$}
}
\startdata
ngc1023-119 & 2:40:35.8 & +38:58:00.9 & 19.022 & 0.001 & 0.987 & 0.002 & 0.616 & 0.001 \\
ngc1023-133 & 2:40:09.7 & +38:58:07.2 & 19.100 & 0.001 & 0.800 & 0.002 & 0.466 & 0.001 \\
ngc1023-450 & 2:40:37.2 & +39:02:06.9 & 19.116 & 0.001 & 0.814 & 0.002 & 0.484 & 0.001 \\
ngc1023-525 & 2:40:20.5 & +39:02:47.6 & 18.903 & 0.001 & 0.573 & 0.002 & 0.347 & 0.001 \\
ngc1023-631 & 2:40:15.8 & +39:03:30.2 & 19.289 & 0.001 & 0.754 & 0.002 & 0.406 & 0.001 \\
ngc1023-813 & 2:40:27.8 & +39:04:40.7 & 19.155 & 0.001 & 0.738 & 0.002 & 0.520 & 0.001 \\
ngc1023-870 & 2:40:28.2 & +39:05:17.6 & 19.056 & 0.001 & 0.733 & 0.002 & 0.430 & 0.001 \\
\enddata
\protect\label{table:n1023 bright objects}
\end{deluxetable}

\begin{deluxetable}{lccc}
\tablecaption{50\% Completeness Limits in Final, Combined Images}
\tablehead{\colhead{} &\multicolumn{3}{c}{Galaxy} \\
\colhead{Filter} & \colhead{NGC~1023} & \colhead{NGC~1055} & \colhead{NGC~7332/7339} } 
\startdata
$B$ & 24.9 & 24.5 & 25.8\\
$V$ & 25.4 & 24.0 & 25.1\\
$R$ & 24.9 & 23.8 & 25.2\\
\enddata
\protect\label{table:completeness}
\end{deluxetable}

\begin{deluxetable}{lcrr}
\tablecaption{GCLF Peak Magnitude and Fractional Coverage}
\tablehead{\colhead{Galaxy} &\colhead{$m_V$(peak)} & \colhead{Mean Coverage}}
\startdata
NGC~1023 & 22.96 & $0.57 \pm 0.03$\\
NGC~1055 & 23.73 & $0.33 \pm 0.01$\\
NGC~7332 & 24.48 & $0.36 \pm 0.03$\\
NGC~7339 & 24.42 & $0.33 \pm 0.02$\\
\enddata
\protect\label{table:gclf}
\end{deluxetable}

\begin{deluxetable}{lrc}
\tablecaption{Corrected Radial Profile of the GC System of NGC~1023}
\tablehead{\colhead{Radius} & \colhead{Surface Density} & \colhead{Source}\\
\colhead{(arcmin)} & \colhead{(arcmin$^{-2}$)} & \colhead{}}
\startdata
0.38 & $62.43 \pm 17.01$ & HST \\
0.59 & $40.62 \pm 8.07$ & HST \\
0.76 & $14.42 \pm 2.97$ & WIYN\\
0.83 & $29.02 \pm 5.64$ & HST \\
1.08 & $13.17 \pm 3.52$ & HST \\
1.33 & $14.22 \pm 3.49$ & HST \\
1.50 & $8.94 \pm 1.59$ & WIYN\\
1.56 & $9.15 \pm 3.03$ & HST \\
1.82 & $10.06 \pm 4.02$ & HST \\
2.07 & $7.73 \pm 4.03$ & HST \\
2.26 & $5.91 \pm 1.27$ & WIYN\\
2.32 & $9.48 \pm 4.80$ & HST \\
2.51 & $7.99 \pm 9.20$ & HST \\
3.09 & $4.15 \pm 1.03$ & WIYN\\
3.88 & $0.93 \pm 0.60$ & WIYN\\
4.65 & $0.43 \pm 0.53$ & WIYN\\
5.47 & $2.07 \pm 0.84$ & WIYN\\
6.27 & $0.58 \pm 0.65$ & WIYN\\
7.04 & $-0.21 \pm 0.50$ & WIYN\\
7.79 & $0.42 \pm 1.01$ & WIYN\\
\enddata
\tablecomments{ Negative surface densities can occur due to
  the application of a subtractive contamination correction.} 
\protect\label{table:1023radprof}
\end{deluxetable}

\begin{deluxetable}{lr}
\tablecaption{Corrected Radial Profile of the GC System of NGC~1055}
\tablehead{\colhead{Radius} & \colhead{Surface Density} \\
\colhead{(arcmin)} & \colhead{(arcmin$^{-2}$)}}
\startdata
0.86 & $11.62 \pm 4.04$ \\
1.60 & $3.90 \pm 1.49$ \\
2.37 & $0.54 \pm 0.56$ \\
3.16 & $1.69 \pm 0.70$ \\
3.94 & $0.72 \pm 0.53$ \\
4.72 & $1.01 \pm 0.59$ \\
5.52 & $-0.05 \pm 0.38$ \\
6.27 & $0.12 \pm 0.46$ \\
\enddata
\tablecomments{ Negative surface densities can occur due to
  the application of a subtractive contamination correction.} 
\protect\label{table:1055radprof}
\end{deluxetable}

\begin{deluxetable}{lrc}
\tablecaption{Corrected Radial Profile of the GC System of NGC~7332}
\tablehead{\colhead{Radius} & \colhead{Surface Density} & \colhead{Source}\\
\colhead{(arcmin)} & \colhead{(arcmin$^{-2}$)} & \colhead{}}
\startdata
0.38 & $54.16 \pm 1.73$ & Keck \\
0.53 & $42.04 \pm 1.62$ & Keck \\
0.53 & $26.16 \pm 7.98$ & WIYN\\
0.75 & $12.99 \pm 1.11$ & Keck \\
0.97 & $11.79 \pm 3.55$ & WIYN\\
1.02 & $8.78 \pm 0.94$ & Keck \\
1.46 & $8.80 \pm 2.50$ & WIYN\\
1.47 & $11.06 \pm 1.04$ & Keck \\
1.68 & $6.22 \pm 0.79$ & Keck \\
1.75 & $1.24 \pm 0.09$ & Keck \\
1.85 & $2.59 \pm 0.41$ & Keck \\
1.95 & $0.72 \pm 1.12$ & WIYN\\
2.46 & $0.56 \pm 1.04$ & WIYN\\
2.94 & $-0.39 \pm 0.50$ & WIYN\\
3.44 & $0.20 \pm 0.99$ & WIYN\\
3.95 & $-0.10 \pm 0.89$ & WIYN\\
4.45 & $0.90 \pm 1.21$ & WIYN\\
4.95 & $-0.13 \pm 0.84$ & WIYN\\
\enddata
\tablecomments{ Negative surface densities can occur due to
  the application of a subtractive contamination correction.} 
\protect\label{table:7332radprof}
\end{deluxetable}

\begin{deluxetable}{lr}
\tablecaption{Corrected Radial Profile of the GC System of NGC~7339}
\tablehead{\colhead{Radius} & \colhead{Surface Density} \\
\colhead{(arcmin)} & \colhead{(arcmin$^{-2}$)}}
\startdata
0.32 & $ 39.48 \pm  23.81  $ \\
0.57 & $ 16.88 \pm  8.34 $ \\
0.86 & $ 0.47 \pm  2.25 $ \\
1.15 & $ 3.35 \pm  2.96 $ \\
1.46 & $ 1.01 \pm  1.97 $ \\
1.76 & $ 0.33 \pm  1.49 $ \\
2.05 & $ 2.55 \pm  1.94 $ \\
2.35 & $ 1.33 \pm  1.56 $ \\
2.65 & $ 0.53 \pm  1.34 $ \\
2.95 & $ 1.30 \pm  1.55 $ \\
3.25 & $ 1.95 \pm  1.67 $ \\
3.55 & $ -0.57 \pm  0.22 $ \\
3.85 & $ 0.06 \pm  1.30 $ \\
\enddata
\tablecomments{ Negative surface densities can occur due to
  the application of a subtractive contamination correction.} 
\protect\label{table:7339radprof}
\end{deluxetable}

\begin{deluxetable}{lrrrrrr}
\tablecaption{Coefficients from Fitting GC System Radial Profile Data}
\tablehead{\colhead{} & \multicolumn{3}{c}{de~Vaucouleurs Law} &\multicolumn{3}{c}{Power Law} \\
\colhead{Galaxy} & \colhead{a0} & \colhead{a1} & \colhead{$\chi^2 / \nu $} & \colhead{a0} & \colhead{a1} & \colhead{$\chi^2 / \nu $}  }
\startdata
NGC~1023 & $3.41 \pm 0.16$ & $-2.15 \pm 0.15$ & 1.59 & $1.23 \pm 0.03$ & $-1.36 \pm 0.09$ & 1.62\\
NGC~1055 & $3.24 \pm 0.45$ & $-2.32 \pm 0.37$ & 1.27 & $0.93 \pm 0.10$ & $-1.60 \pm 0.25$ & 1.16\\
NGC~7332 & $2.89 \pm 0.22$ & $-1.78 \pm 0.20$ & 1.89 & $1.09 \pm 0.05$ & $-1.24 \pm 0.14$ & 1.63\\
NGC~7339 & $3.43 \pm 0.44$ & $-2.60 \pm 0.41$ & 0.98 & $0.77 \pm 0.09$ & $-1.55 \pm 0.24$ & 0.85\\
\enddata
\protect\label{table:fitting}
\end{deluxetable}

\begin{deluxetable}{lccrcrrc}
\tablecaption{Total $N_{\rm GC}$ and GC Specific Frequencies of Giant Galaxies
with Wide-Field Multi-Color CCD Imaging}
\tablehead{\colhead{Galaxy} & \colhead{Type} & \colhead{$M_V$} & \colhead{$N_{GC}$} & \colhead{$S_N$} & \colhead{$T$} & \colhead{$T_{\rm blue}$} &\colhead{Ref.} } 
\startdata
NGC~1023 & SB0 & $-$21.2 & $490 \pm 30$ & $1.7 \pm 0.3$ & $2.7 \pm 0.4$ & $1.5 \pm 0.2$ & 14 \\
NGC~7332 & S0pec & $-$20.7 & $175 \pm 15$ & $0.9 \pm 0.3$ & $1.4 \pm 0.4$ & $0.8 \pm 0.3$ & 14 \\
NGC~1055 & SBb & $-$21.0 & $210 \pm 40$ & $0.9 \pm 0.2$ & $1.6 \pm 0.5$ & $0.9 \pm 0.3$ & 14 \\
NGC~7339 & Sbc & $-$20.4 & $75 \pm 10$ & $0.5 \pm 0.2$ & $1.5 \pm 0.5$ & $0.9 \pm 0.3$ & 14 \\
NGC~1052 & E4 & $-$21.0 & $400 \pm 45$ & $1.6 \pm 0.3$ & $1.9 \pm 0.5$ & $0.9 \pm 0.3$ & 3 \\
NGC~3379 & E1 & $-$20.9 & $270 \pm 30$ & $1.2 \pm 0.3$ & $1.4 \pm 0.4$ & $1.0 \pm 0.2$ & 11 \\
NGC~4374 & E1 & $-$22.1 & $1775 \pm 380$ & $2.7 \pm 0.6$ & $3.1 \pm 0.7$ & $1.9 \pm 0.4$ & 4 \\
NGC~4406 & E3 & $-$22.3 & $2900 \pm 400$ & $3.5 \pm 0.5$ & $4.1 \pm 0.6$ & $2.5 \pm 0.4$ & 11 \\
NGC~4472 & E2 & $-$23.1 & $5870 \pm 680$ & $3.6 \pm 0.6$ & $4.2 \pm 0.6$ & $2.6 \pm 0.4$ & 9 \\
NGC~5128 & Epec & $-$22.1 & $1550 \pm 390$ & $2.2 \pm 0.6$ & $2.6 \pm 0.7$ & $1.4 \pm 0.3$ & 7 \\
NGC~3384 & SB0 & $-$20.5 & $120 \pm 30$ & $0.8 \pm 0.2$ & $1.2 \pm 0.4$ & $0.7 \pm 0.2$ & 6 \\
NGC~4594 & S0/Sa & $-$22.4 & $1890 \pm 200$ & $2.1 \pm 0.3$ & $3.2 \pm 0.5$ & $2.0 \pm 0.3$ & 11 \\
NGC~7457 & S0 & $-$19.5 & $210 \pm 30$ & $3.1 \pm 0.7$ & $4.8 \pm 1.1$ & $2.8 \pm 0.6$ & 5 \\
Milky~Way & Sbc & $-$21.3 & $160 \pm 20$ & $0.5 \pm 0.1$ & $1.1 \pm 0.1$ & $0.8 \pm 0.1$ & 1 \\
M31 & Sb & $-$21.8 & $450 \pm 100$ & $0.9 \pm 0.2$ & $1.6 \pm 0.4$ & $1.2 \pm 0.3$ & 1,2,8 \\
NGC~891 & Sb & $-$20.8 & $70 \pm 20$ & $0.3 \pm 0.1$ & $0.6 \pm 0.3$ & $0.3 \pm 0.2$ & 13\\
NGC~2683 & Sb & $-$20.5 & $120 \pm 40$ & $0.8 \pm 0.4$ & $1.4 \pm 0.7$ & $0.9 \pm 0.5$ & 12 \\
NGC~3556 & Sc & $-$21.2 & $290 \pm 80$ & $0.9 \pm 0.4$ & $2.2 \pm 0.9$ & $1.4 \pm 0.5$ & 12 \\
NGC~4013 & Sb & $-$20.4 & $140 \pm 20$ & $1.0 \pm 0.2$ & $1.9 \pm 0.5$ & $1.4 \pm 0.3$ & 13 \\
NGC~4157 & Sb & $-$20.4 & $80 \pm 20$ & $0.6 \pm 0.3$ & $1.1 \pm 0.6$ & $0.6 \pm 0.4$ & 12 \\
NGC~7331 & Sb & $-$21.7 & $210 \pm 130$ & $0.5 \pm 0.4$ & $0.9 \pm 0.7$ & $0.4 \pm 0.3$ & 12 \\
NGC~7814 & Sab & $-$20.4 & $165 \pm 25$ & $1.3 \pm 0.4$ & $2.2 \pm 0.8$ & $1.3 \pm 0.5$ & 10 \\
\enddata
\tablerefs{(1) \citealp{az98}; (2) \citealp{barmby00}; (3) \citealp{forbes01}; (4) \citealp{gr04}; 
(5) \citealp{h11}; (6) Hargis \& Rhode 2012a, in preparation; (7) \citealp{harris06}; (8) \citealp{perrett02};
(9) \citealp{rz01}; (10) \citealp{rz03}; (11) \citealp{rz04}; (12) \citealp{rzk07}; 
(13) \citealp{rwy10}; (14) this paper. }
\protect\label{table:results}
\end{deluxetable}

\end{document}